\documentclass[lettersize,journal]{IEEEtran}
\usepackage{graphicx}%
\usepackage{multirow}%
\usepackage{amsmath,amssymb,amsfonts}%
\usepackage{amsthm}
\usepackage{mathrsfs}%
\usepackage{xcolor}%
\usepackage{textcomp}%
\usepackage{manyfoot}%
\usepackage{booktabs}%
\usepackage{algorithm}%
\usepackage{algorithmicx}%
\usepackage{algpseudocode}%
\usepackage{listings}%
\usepackage{lipsum}
\usepackage{subcaption}
\usepackage{makecell}
\usepackage{geometry}
\usepackage{float}
\usepackage[noadjust]{cite}
\usepackage{url}
\usepackage{hyperref}
\usepackage[nameinlink,capitalise]{cleveref}
\hypersetup{colorlinks}
\hypersetup{
    citecolor=cyan,
    linkcolor=blue,
    filecolor=green,      
    urlcolor=cyan,
}

\newcommand{\doi}[1]{doi.~\url{https://doi.org/#1}}

\usepackage[utf8]{inputenc}
\usepackage{fourier} 
\usepackage{array}
\usepackage{makecell}

\begin{document}

\title{Optimized Single-Core PCF-Based SPR Biosensor for High-Performance Early-Stage Multi-Cancer Detection}

\author{Tonmoy Malakar$^{1}$, Miss Nourin Nurain Amina$^{2*}$, Nazmus Shakib Lalin$^2$, Zarin Tasnim Nijhum$^2$\\~\IEEEmembership{$^{[1]}$ Electrical and Computer Engineering, The University of Oklahoma, Norman, USA\\
$^{[2]}$ Physics Discipline, Khulna University, Khulna, 9208, Bangladesh\\
nurainnourin@gmail.com}}



\maketitle

\begin{abstract}
In this study, we present a highly sensitive Surface Plasmon Resonance (SPR)-based biosensor integrated with a circular-lattice Photonic Crystal Fiber (PCF) for early-stage cancer detection. The proposed sensor leverages the synergy between SPR and PCF technologies to overcome the bulkiness and limited sensitivity of traditional SPR systems. A thin gold (Au) layer, responsible for plasmon excitation, is deposited on the fiber structure, while a nanolayer of vanadium pentoxide $(V_2O_5)$ is introduced to enhance adhesion between the gold and the silica background, improving structural stability and field confinement. The sensor is designed to detect refractive index (RI) variations in biological analytes, specifically targeting cancerous cells from skin, blood, and adrenal gland tissues. The optical characteristics and performance of the sensor were thoroughly analyzed using the Finite Element Method (FEM) in COMSOL Multiphysics 6.1, allowing for precise simulation and optimization. The sensor demonstrates high sensitivity within the RI range of 1.360 –1.395, corresponding to the RI values of the target cancer cells. Remarkable wavelength sensitivities of 21,250 $nm/RIU$, 53,571 $nm/RIU$, and 103,571 $nm/RIU$ were achieved for skin, blood, and adrenal gland cancers, respectively. In addition, a maximum figure of merit (FOM) of 306.424 $ RIU^{-1}$ and a spectral resolution (SR) of $9.57 \times10^{-7} RIU$ further affirm the sensor’s exceptional detection capabilities. These findings indicate the proposed SPR-PCF sensor’s strong potential for real-time, label-free biosensing applications, particularly in precise and early cancer diagnostics.
\end{abstract}


\section{Introduction}
\IEEEPARstart{C}{ancer} remains one of the most critical global public health challenges, currently ranking as the second leading cause of mortality worldwide, surpassing even heart disease in some regions \cite{siegel2017colorectal}. According to the World Health Organization (WHO), cancer is responsible for approximately one in six deaths globally, with an estimated 19.3 million new cases and 10 million deaths recorded in 2020 alone \cite{singh2019nanotechnology, sung2021global}. Characterized by the uncontrolled proliferation of abnormal cells, cancer disrupts essential biological functions and often leads to fatal outcomes \cite{siegel2017colorectal}. This unchecked cellular growth can be triggered by a range of factors, including exposure to toxic substances, radiation, fungal agents, poor dietary habits, obesity, excessive alcohol consumption, physical inactivity, and nutritional deficiencies \cite{siegel2017colorectal}. Among the various forms of cancer, skin cancer and blood cancer are among the most frequently diagnosed types, with skin cancer showing a rising global incidence \cite{siegel2017colorectal}. Despite significant advancements in medicine, a definitive cure for cancer remains elusive \cite{siegel2017colorectal}. Therefore, early and accurate detection is paramount, as it substantially increases the chances of successful treatment and patient survival \cite{liang2007determining, sharma2015photonic}. Traditional diagnostic methods such as biopsies, physical exams, endoscopy, and imaging techniques like MRI and X-rays are often expensive and rely on bulky, complex equipment \cite{sharma2015photonic}. Consequently, there is an urgent demand for rapid, cost-effective, and highly sensitive diagnostic technologies that can facilitate early cancer detection \cite{sharma2015photonic}.

Optical biosensors have emerged as a leading solution for early disease detection due to their high sensitivity, compactness, rapid response, and low-cost fabrication. These advantages make them ideal for real-time biomedical applications such as cancer diagnosis. Various optical techniques, including fluorescence imaging, Raman spectroscopy, photothermal methods, terahertz (THz) spectroscopy, microfluidic devices, and fiber-optic sensors, have been explored for this purpose \cite{zaytsev2015vivo, hajba2014circulating, liu2014surface, li2013detection}. Among these, Photonic Crystal Fiber (PCF)-based sensors stand out for their superior light-guiding properties and design flexibility. Unlike conventional fibers, PCFs feature periodic air-hole structures that provide enhanced control over dispersion, birefringence, and mode confinement. These properties allow PCFs to be tailored for ultra-sensitive detection of biological analytes. As a result, PCFs are widely used in sensing applications such as temperature, pressure, glucose, pH, DNA, and cancer biomarker detection. Their microstructured architecture enables strong interaction between light and analyte, leading to improved sensitivity and precision. With their compact design, low optical loss, and ability to perform label-free detection, PCF-based biosensors offer a powerful platform for next-generation, non-invasive cancer diagnostics.

The development of biosensors, particularly surface plasmon resonance (SPR)-based platforms, has significantly advanced over recent decades. The earliest SPR biosensor, designed by Clark in 1985, was utilized for glucose detection in blood \cite{hasan2023indium}. Since then, substantial progress has been made in applying SPR and photonic crystal fiber (PCF)-based sensors for biomedical diagnostics, particularly in cancer detection. Ahmet Yasli investigated SPR-based PCF biosensors \cite{yasli2021cancer}, while other studies proposed optimized single-core \cite{jabin2019surface} and dual-core PCF designs \cite{ayyanar2018photonic} for detecting various cancerous cells, including those affecting the breast, basal, and cervical tissues. A triple-core PCF biosensor was introduced for differentiating between normal and malignant blood cells based on refractive index (RI) contrast, offering early cancer detection capability\cite{gong2024lab}. Bulbul et al. explored PCF sensors in the terahertz (THz) regime using the finite element method for breast cancer detection \cite{bulbul2020design}. Meanwhile, Rakhshani et al. developed a plasmonic sensor using a 3D nanorod metasurface, proposing a wide-angle absorber for cancer cell identification \cite{rakhshani2021wide}. Recent innovations have focused on enhancing sensitivity and selectivity through novel material integration and hybrid configurations. For instance, Hemanth et al. introduced a 2D photonic crystal-based cervical cancer sensor achieving a peak sensitivity of 143 $nm/RIU$ \cite{kumar20202d}. Karki et al. made significant contributions to SPR sensor design by incorporating various layered materials such as black phosphorus (BP) \cite{karki2025tuning}, platinum diselenide $(PtSe_2)$ and graphene \cite{karki2024platinum}, zinc sulfide (ZnS), silicon dioxide (SiO\textsubscript{2}) \cite{karki2022zinc}, and MXene \cite{karki2024gold} to enhance performance. Their sensors demonstrated high sensitivity (up to $374.31^\circ/\text{RIU}$) and figure of merit (FOM), though challenges in fabrication complexity and material stability were noted. Further, Karki and collaborators reviewed state-of-the-art SPR biosensing technologies for cancer detection and discussed avenues for improving sensitivity, specificity, and real-world applicability \cite{karki2022advances}. Kumar et al. developed a microfiber interferometer biosensor for ultra-low concentration detection \cite{kumar2019ultrasensitive}, albeit without thoroughly addressing integration with current diagnostic frameworks. Pal et al. \cite{pal2024highly} and others validated long-range SPR sensors for blood-based biomarker detection. More recently, simulation-based designs have achieved remarkable sensitivities. Jabin et al. developed a D-shaped breast cancer sensor with peak sensitivity reaching 18,000 $nm/RIU$, surpassing most prior designs focused predominantly on basal cells \cite{jabin2019surface}.

All fiber-based surface plasmon resonance (SPR) biosensors rely on the phase-matched coupling between the core-guided mode and the surface plasmon polariton (SPP) mode at the metal-dielectric interface. This coupling facilitates precise detection of refractive index (RI) variations in cellular samples by monitoring shifts in the resonance wavelength. When resonance conditions are satisfied, a distinct Gaussian-shaped peak emerges in the confinement loss spectrum, corresponding to the analyte’s specific RI\cite{malakar2025performance}. The excitation of surface plasmons is highly dependent on the properties of the metal layer, which serves as the plasmonic medium. Common plasmonic materials include silver (Ag), copper (Cu), and gold (Au), among which gold is preferred due to its superior chemical stability, resistance to oxidation, biocompatibility, and ability to produce sharp and well-defined resonance peaks \cite{sultana2021highly}. Photonic crystal fibers (PCFs) differ fundamentally from conventional optical fibers by incorporating a periodic array of air holes in the cladding, which induces spatial modulation of the refractive index and enhances light confinement within the core. These air holes are typically arranged in circular, hexagonal, or hybrid geometries, and can be engineered to tailor the modal properties of the fiber. In SPR-PCF biosensor designs, a section of the fiber is selectively infiltrated with the analyte, while a nanoscale metal film is strategically deposited at the core-cladding boundary or within inner air holes to facilitate plasmon excitation \cite{haque2021highly}.

Traditional cancer detection methods, especially for blood-related cancers, are often expensive, invasive, and lack sufficient sensitivity. In this research, a single-core photonic crystal fiber based SPR biosensor is proposed to detect cancer cells in blood, skin, and adrenal gland tissues. The sensor works by observing shifts in the transmission and loss spectra, which occur due to changes in the refractive index (RI) of different cancer cells. These RI changes cause a resonance shift that is independent of cell size but directly linked to the coupling between the guided light and the biofluid at a specific wavelength. The U-shaped analyte channel enables direct interaction between the analyte and the plasmonic surface, minimizing the need for repeated analyte refilling and enhancing the sensor's suitability for real-time applications. A nanocomposite layer of vanadium pentoxide $(V_2O_5)$ nanoparticles placed beneath a thin gold coating is used to improve plasmon excitation and enhance sensitivity. Key structural parameters, such as air hole diameter, pitch, and gold layer thickness, are optimized using the finite element method (FEM) to achieve high performance. The proposed sensor offers high sensitivity, low confinement loss, and the ability to detect various cancer types quickly and accurately. These features make it a promising tool for practical biomedical sensing and early cancer diagnosis.

In this study, a novel single-core photonic crystal fiber (PCF) design is proposed for the rapid and sensitive detection of various malignant cells, including basal skin cancer cells, Jurkat blood cancer cells, and PC12 adrenal gland cancer cells. The sensor demonstrates a notable blue shift in its spectral response, enhancing its effectiveness for cancer detection. It achieves a maximum wavelength sensitivity (WS) of 103,571 $nm/RIU$. The sensor’s performance is rigorously evaluated using the finite element method (FEM), which enables accurate modeling of the complex structure, simulation of electromagnetic field behavior, and optimization of key parameters for maximum sensitivity. Comparative analysis between cancerous and normal cells shows clear sensitivity differences, validating the biosensor’s diagnostic potential. This design offers strong advantages in detecting cancer biomarkers at low concentrations, making it highly suitable for early-stage cancer diagnosis. Additionally, the sensor has potential applications in broader healthcare diagnostics, including disease monitoring and assessment of treatment effectiveness over time. Overall, the proposed SPR-PCF biosensor represents a promising tool for advancing real-time, non-invasive cancer detection.

\section{Design Strategy and Theoretical Modeling}

\begin{figure*}[h]
    \centering
    \subfloat[]{
        \includegraphics[width=.55\linewidth]{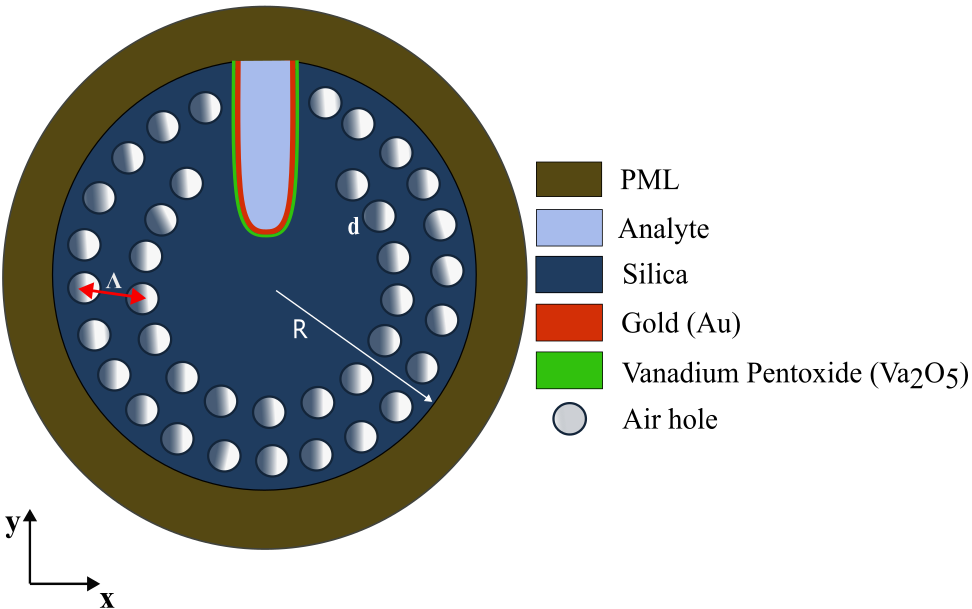}
    }
    \subfloat[]{
        \raisebox{25mm}{ 
\includegraphics[width=.40\linewidth]{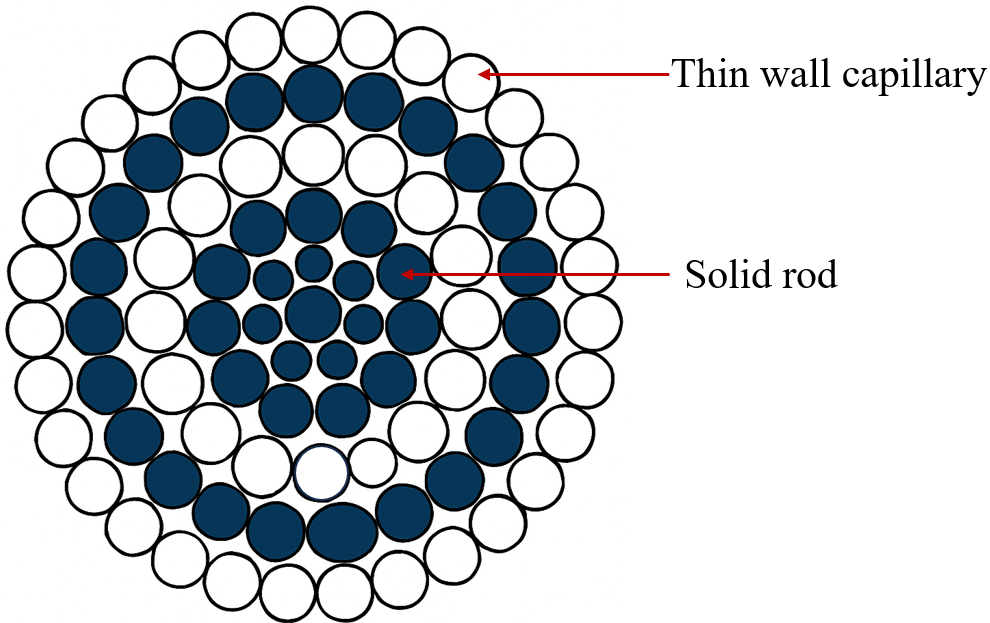}
        }
    }\\
    \caption{\textbf{(a)} Cross-sectional view of the proposed sensor design showing the material layers and structural components, \textbf{(b)} schematic layout of the glass capillary and rod assembly used in the preform fabrication}
    \label{Fig:1}
\end{figure*}

\noindent The structural configuration of the proposed SPR-based single-core photonic crystal fiber (PCF) biosensor, along with its x–y cross-sectional representation, is illustrated in the 2D schematic shown in \cref{Fig:1}.

The proposed photonic crystal fiber (PCF)-based biosensor features a carefully engineered structure optimized for surface plasmon resonance (SPR) excitation, as illustrated in \cref{Fig:1}. The cross-sectional design includes two concentric rings of uniformly sized air holes, each with diameter $d$, arranged in a circular lattice. The distance between the centers of adjacent air holes, known as the pitch $(\Lambda)$, is kept constant throughout the structure.

To form the U-shaped analyte channel, specific air holes from the innermost and second ring are intentionally omitted. This creates a region that supports strong coupling between the fundamental core-guided mode and the surface plasmon polariton (SPP) mode, which is essential for efficient SPR excitation. Notably, no air holes are placed between the fiber core and the metal–dielectric interface, providing an expanded interaction region and facilitating smooth power transfer from the core mode to the SPP mode.

A thin gold $(Au)$ layer is deposited on the outer surface of the analyte channel, serving as the plasmonic medium for SPR excitation. To further enhance performance, a vanadium pentoxide $(V_2O_5)$ layer is introduced beneath the gold layer. $V_2O_5$ not only increases sensitivity by amplifying the interaction between the evanescent field and the analyte, but also provides chemical stability due to its fully oxidized state. Gold, being chemically inert and resistant to oxidation, acts as a robust protective barrier over $V_2O_5$ and the sensor's structural components. If necessary, environmental isolation or additional protective coatings may be applied to further preserve performance.

The fabrication of this structure can be achieved using the stack-and-draw technique \cite{rifat2017highly}, a well-established method for PCF production. As shown in \cref{Fig:1}b, the preform is assembled by stacking thin-walled glass capillaries and solid rods in a pattern that replicates the desired fiber geometry. The U-shaped channel is then created using the polishing method described in \cite{wu2018ultrahigh}. Plasmonic materials, including gold and $V_2O_5$, are subsequently deposited onto the polished surface using chemical deposition techniques \cite{chen2018surface}.

The sensor’s performance is evaluated using the full-vector finite element method (FEM) in COMSOL Multiphysics (version 6.1). To minimize boundary reflections and improve radiative absorption, a perfectly matched layer (PML) is applied at the simulation domain boundaries. A finer meshing strategy is employed to enhance numerical accuracy. The main structural parameters optimized in the design include the air hole radius $(r)$, the pitch $(\Lambda)$, the gold layer thickness $(t_{Au})$, and the microchannel opening width, which is set to 6 $\mu m$.
The values of the biosensor's geometrical parameters are listed in \ref{tab:Table-1}.

\begin{table}[h]
    \caption{Parameter values of the proposed biosensor}
    \label{tab:Table-1}
    \centering
    \renewcommand{\arraystretch}{1.5}
    \begin{tabular}{c c c c}
    \hline
    \thead{Cladding radius} & \thead{Air hole radius} & \thead{Pitch} &  \thead{Gold layer\\ thickness} \\
    R = 33 $\mu m$ & r = 2.3 $\mu m$ &  $\Lambda=5.7$ $\mu m$ & $t_{Au}=37$ $nm$\\
    \hline
    \thead{$V_2O_5$\\ adhesive thickness} & \thead{Perfectly matched\\layer}\\
    $t_{V_2O_5}=15$ $nm$ & $PML=9$ $\mu m$\\
    \hline
    \end{tabular}
\label{1}
\end{table}
Silica is used as the background material of the proposed biosensor, and its refractive index (RI) is determined using the well-known Sellmeier equation, given by \cite{hassan2020liquid5}:
\begin{equation}\label{Eq:1}
   n_{Si}^2(\lambda)=1+\frac{B_1\lambda^2}{\lambda^{2}-C_1} +\frac{B_2\lambda^2}{\lambda^{2}-C_2}+\frac{B_3\lambda^2}{\lambda^{2}-C_3} 
\end{equation}
where $\lambda$ represents the incident light wavelength in $\mu m$, and $B_n$ and $C_n$ are the Sellmeier coefficients for n = 1, 2, 3. The values of $B_1$, $B_2$, and $B_3$ are 0.6961663, 0.4079426, and 0.8974794, respectively, while $C_1$, $C_2$, and $C_3$ are 0.068404, 0.1162414, and 9.896161 $\mu m$, as given in \cite{ullah2024highly6}.
The frequency-dependent refractive index of the adhesive material, $V_2O_5$, is calculated using the following equation: \cite{atuchin2014effects}:
\begin{equation}\label{Eq:2}
    n_{V_2O_5}=A+\frac{B}{\lambda^2}+\frac{C}{\lambda^4}
\end{equation}
where $\lambda$ is incident wavelength in $\mu m$. A, B, and C are empirical fitting parameters that describe how the refractive index of a transparent material varies with wavelength.
The relative permittivity of gold is typically modeled using the Drude–Lorentz model, given by the following expression:
\begin{equation}\label{Eq:3}
    \epsilon_{Au}=\epsilon_\infty-\frac{\omega_D^2}{\omega(\omega+j\gamma_D)}-\frac{\Delta\epsilon\Omega_L^2}{(\omega^2-\Omega_L^2)+j\Gamma_L\omega}
\end{equation}
where $\epsilon_\infty=5.9673$ is the permittivity at high frequency, $\Delta\epsilon = 1.09$ is weighting factor, and $\omega$ is the angular frequency. The plasma frequency $\omega_D$ and damping frequency $\gamma_D$ are expressed by $\omega_D/{2\pi}= 2113.6$ THz and $\gamma_D/{2\pi}= 15.92$ THz, respectively. Frequency of the Lorentz oscillators and spectral width are given by $\Omega_L/{2\pi}= 650.07$ THz and $\Gamma_L/{2\pi}= 104.86$ THz, respectively \cite{zhu2017surface8}.

\section{Computational Results and Sensor Performance}
\begin{figure*}[h]
    \centering
    \subfloat[]{
        \includegraphics[width=0.25\linewidth]{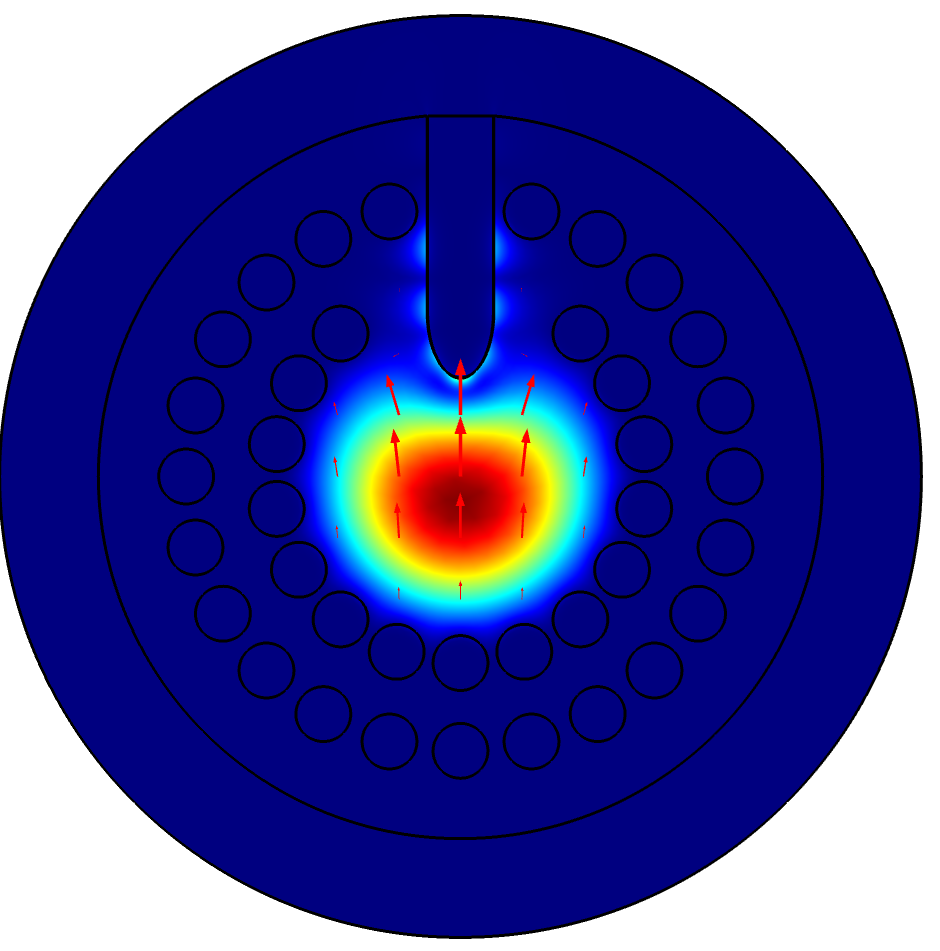}
    }
    \hspace{0.05\linewidth}
    \subfloat[]{
        \includegraphics[width=0.25\linewidth]{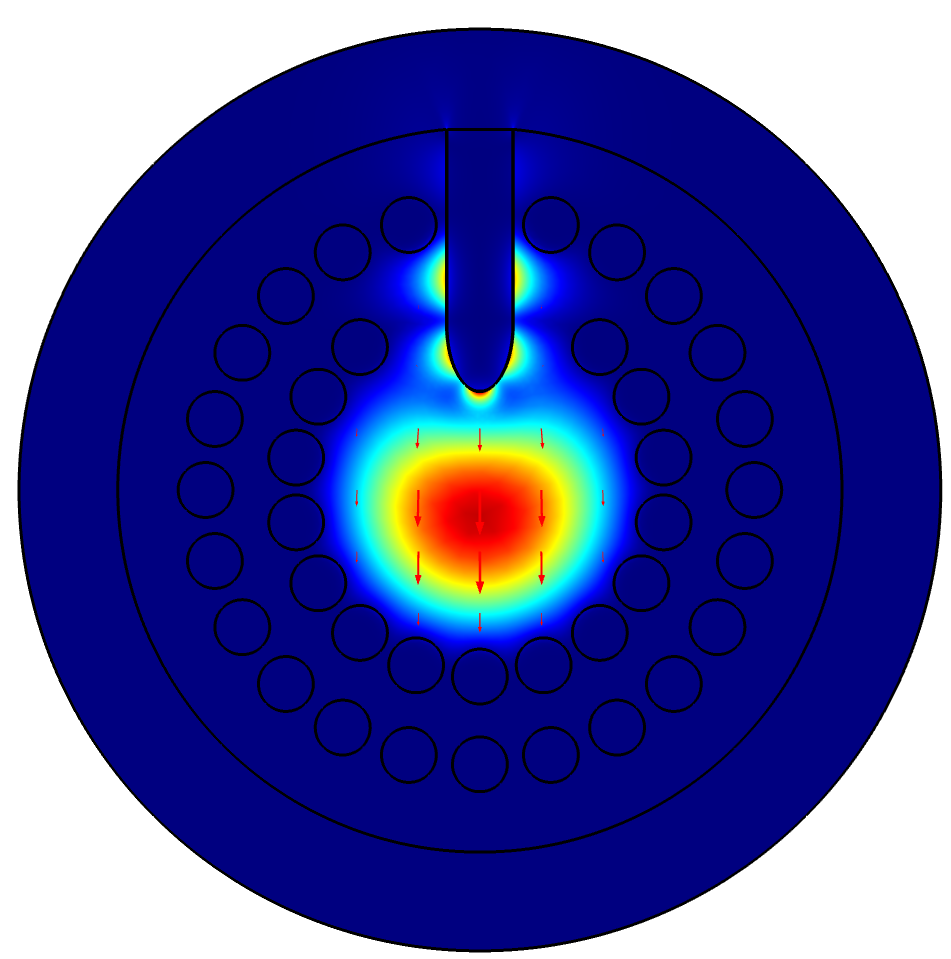}
    }\\[1ex]
    \subfloat[]{
        \includegraphics[width=0.25\linewidth]{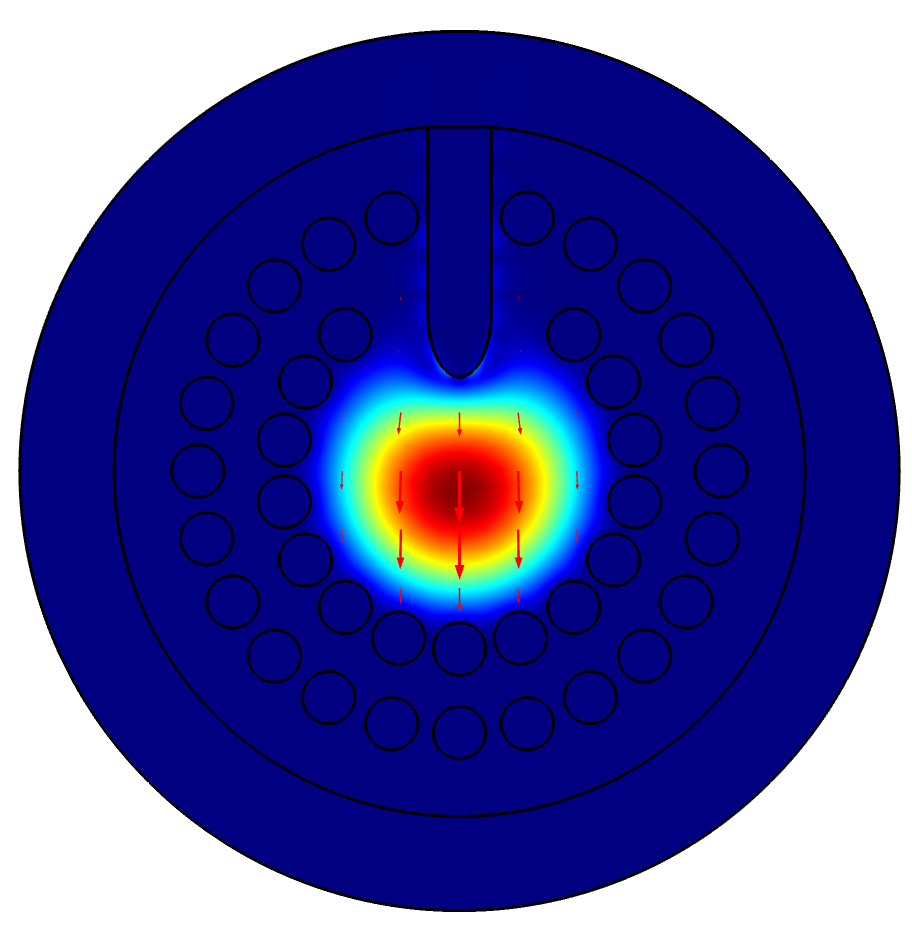}
    }
    \hspace{0.05\linewidth}
    \subfloat[]{
        \includegraphics[width=0.25\linewidth]{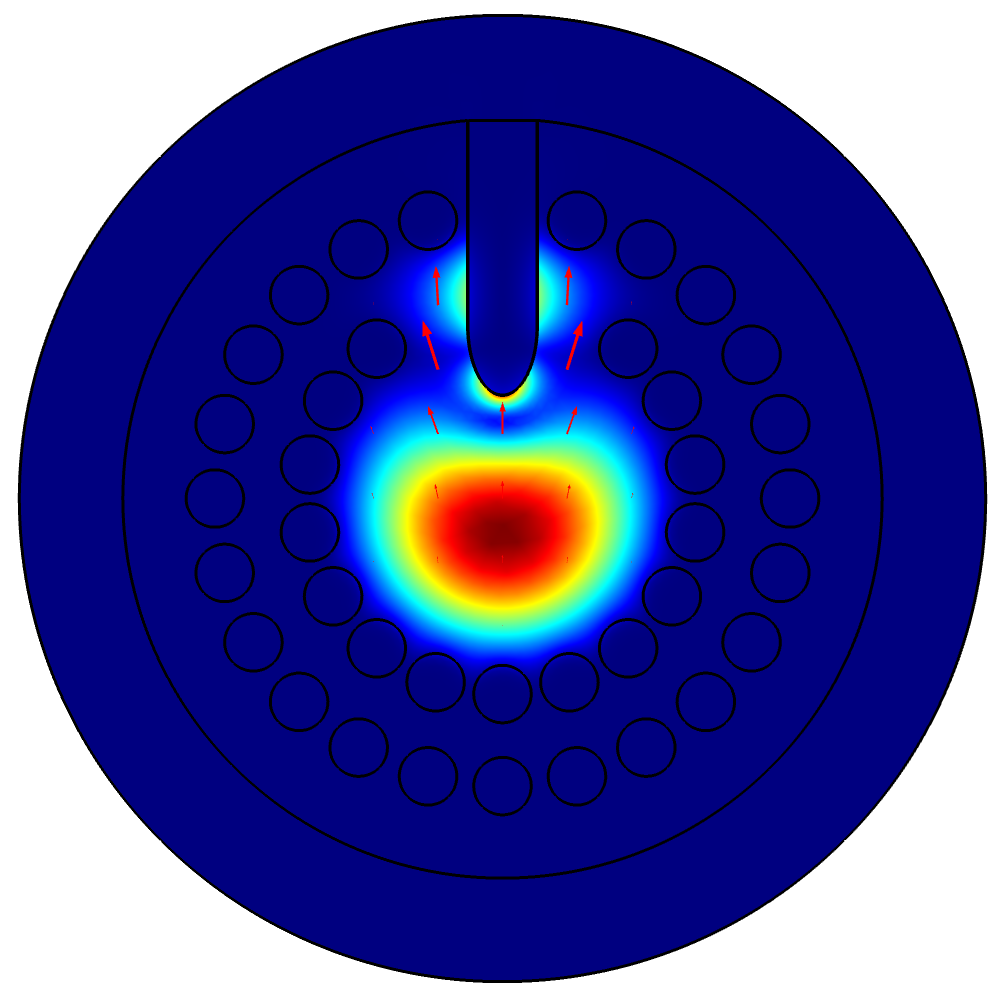}
    }\\[1ex]
    \subfloat[]{
        \includegraphics[width=0.25\linewidth]{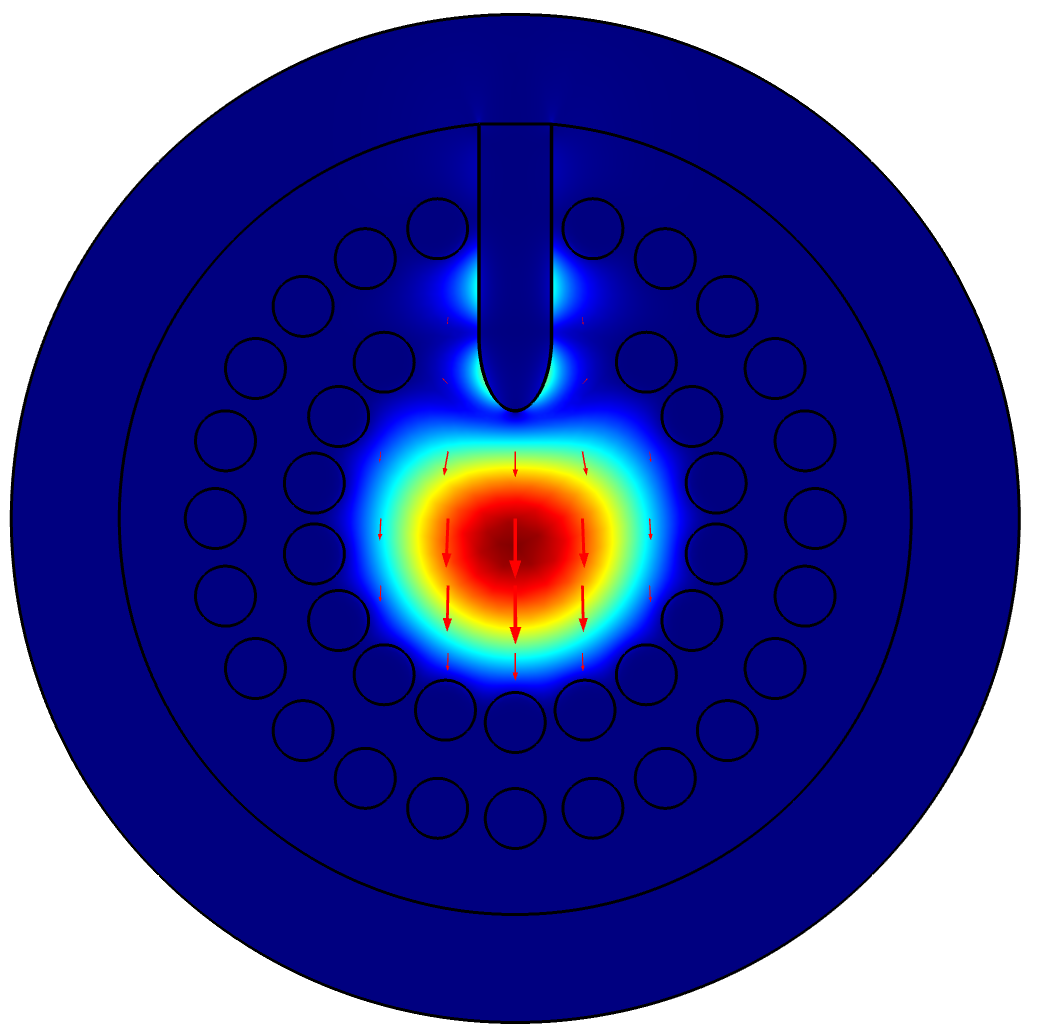}
    }
    \hspace{0.05\linewidth}
    \subfloat[]{
        \includegraphics[width=0.25\linewidth]{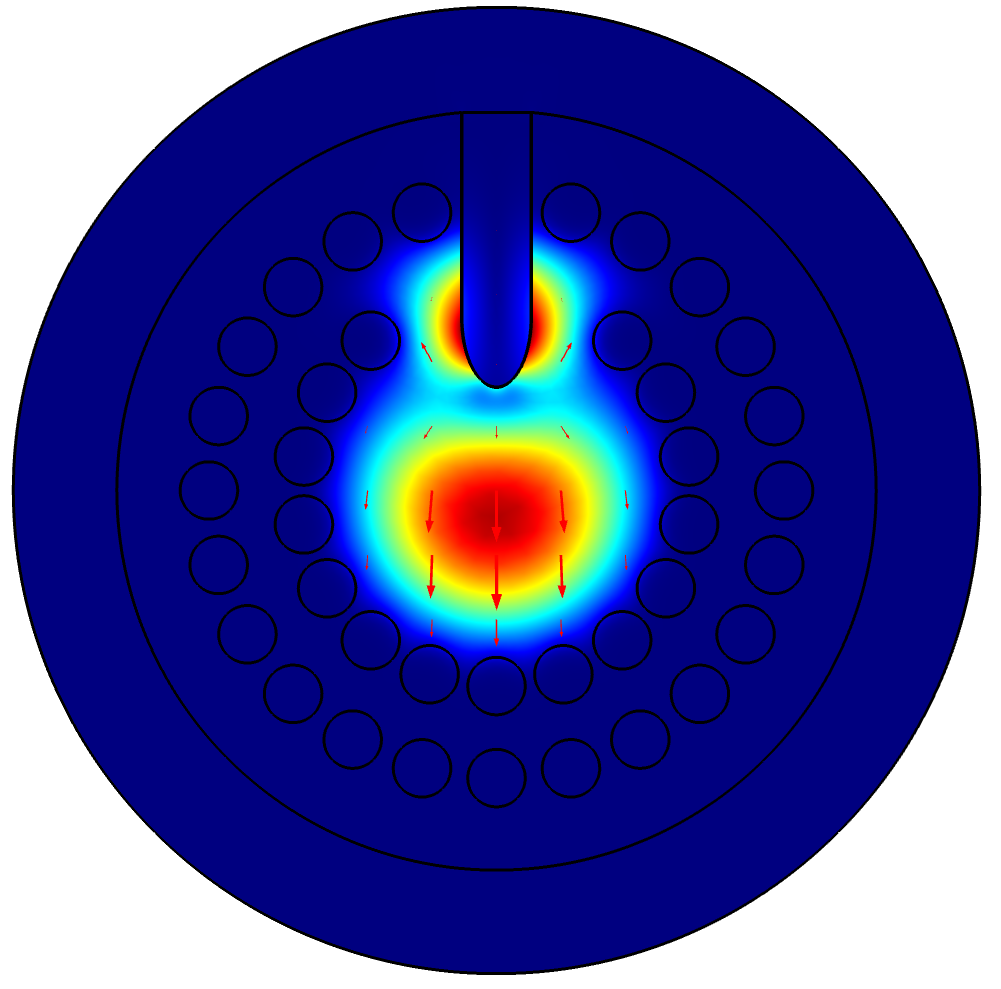}
    }
    
    \caption{Electric field distributions at the surface plasmon resonance (SPR) condition for three biological cell types using the proposed SPR-PCF biosensor. (a) Normal skin cell (RI = 1.36) at 1.925 $\mu m$; (b) Cancerous skin cell (RI = 1.38) at 2.225 $\mu m$; (c) Normal blood cell (RI = 1.376) at 2.1 $\mu m$; (d) Jurkat cell (RI = 1.39) at 2.85 $\mu m$; (e) Normal adrenal gland cell (RI = 1.381) at 2.4 $\mu m$; (f) Adrenocortical carcinoma cell (RI = 1.395) at 3.775 $\mu m$. In all cases, strong field confinement near the metal-dielectric interface and redshifts in the resonance wavelengths demonstrate effective plasmonic coupling and the sensor’s high refractive index sensitivity for reliable cancer cell detection.}
    \label{fig:2}
\end{figure*}

To assess the effectiveness of the proposed SPR-PCF biosensor in discriminating between normal and cancerous cells, comprehensive numerical simulations were performed for three biologically significant cell types: skin cells, blood cells, and adrenal gland cells. Each simulation modeled both healthy and malignant states by assigning appropriate refractive indices (RIs) based on literature values.\\
For skin cell analysis, an RI of 1.36 was used for normal cells and 1.38 for cancerous cells, yielding SPR peaks at 1.925 $\mu m$ and 2.225 $\mu m$, respectively, as shown in \cref{fig:2}(a) and \cref{fig:2}(b). Similarly, in the case of blood cells, a normal cell (RI = 1.376) and a Jurkat leukemia cell (RI = 1.39) produced resonance wavelengths at 2.1 $\mu m$ and 2.85 $\mu m$, respectively, as depicted in \cref{fig:2}(c-d). For adrenal gland cells, a normal cell with an RI of 1.381 and an adrenocortical carcinoma cell with an RI of 1.395 exhibited resonance conditions at 2.4 $\mu m$ and 3.775 $\mu m$, respectively, as illustrated in \cref{fig:2}(e-f). In all cases, the observed redshift in the resonance wavelength with increasing RI reflects a strong and consistent plasmonic response to minute changes in the analyte environment. These shifts are attributed to the phase-matching condition between the core-guided mode and the surface plasmon polariton (SPP) mode, where enhanced electromagnetic field confinement occurs at the metal-dielectric interface. The clear spatial localization of the field near the sensing region further validates the sensor’s ability to support strong mode coupling. Overall, the distinct spectral responses across varying cell types highlight the biosensor's high refractive index sensitivity and its significant potential for label-free, real-time cancer cell detection in biomedical applications.

\subsection{Confinement Loss}
\begin{figure*}[h]
    \centering
    \subfloat[]{
    \includegraphics[width=0.50\linewidth]{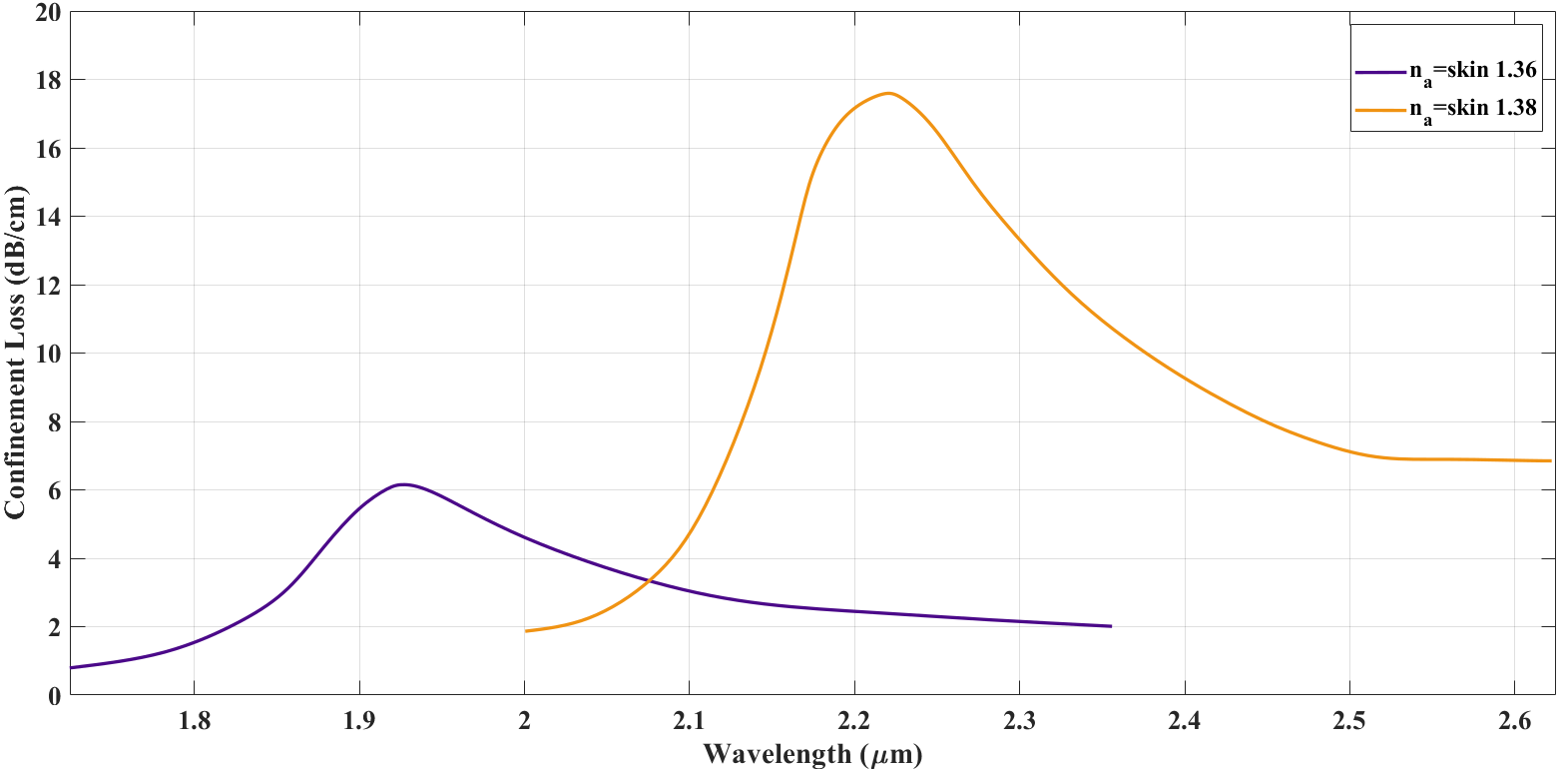}}
    \subfloat[]{
    \includegraphics[width=0.50\linewidth]{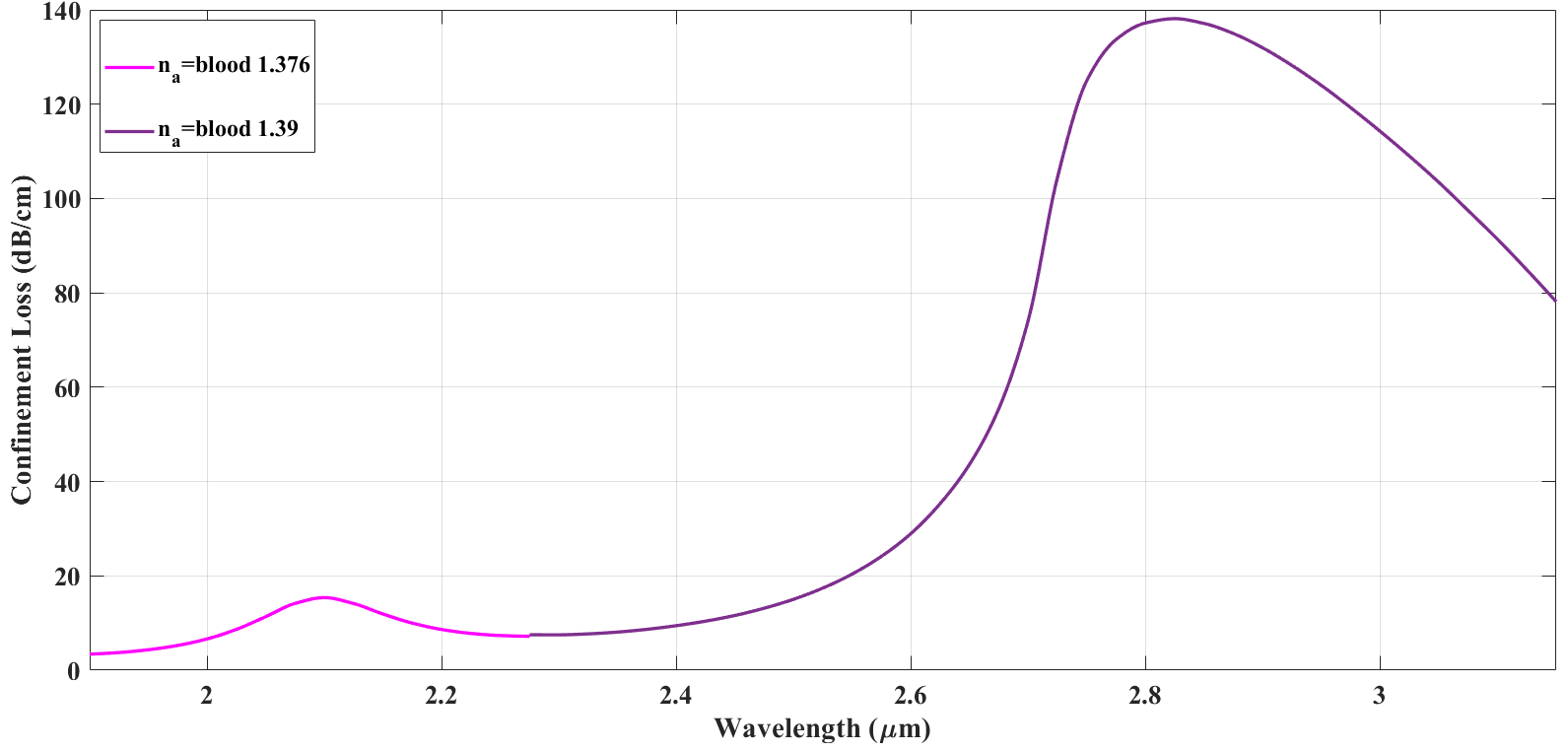}}
    \\[2ex]
    \subfloat[]{
        \includegraphics[width=0.50\linewidth]{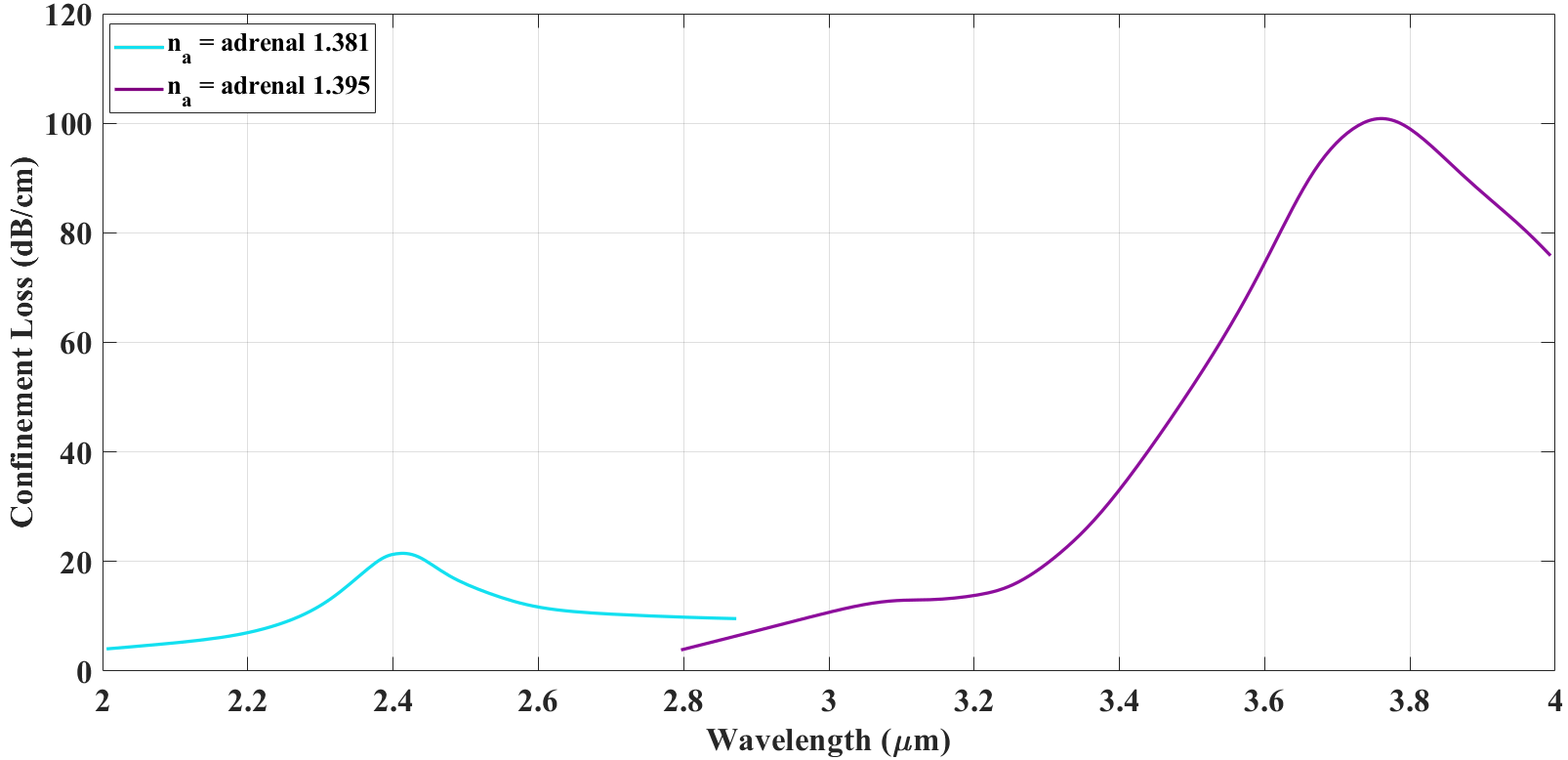}
    }
    \caption{Confinement loss (CL) spectra of the proposed PCF-SPR biosensor for differentiating between normal and cancerous cells: (a) skin cell with refractive indices n = 1.36 (normal) and n = 1.38 (cancerous), (b) blood cell with n = 1.376 (normal) and n = 1.39 (cancerous), and (c) adrenal gland cell with n = 1.381 (normal) and n = 1.395 (cancerous). The distinct peaks in CL indicate enhanced plasmonic interaction in cancerous samples, demonstrating the sensor’s high sensitivity to biomarker-induced refractive index variations.}
    \label{fig:3}
\end{figure*}

Confinement loss is a critical parameter in evaluating the performance of SPR-based photonic crystal fiber (PCF) sensors, as it quantifies the leakage of optical power from the guided mode into the plasmonic layer. It directly reflects the degree of mode coupling between the core-guided light and the surface plasmon polariton (SPP) mode at the metal-dielectric interface. This loss is intrinsically linked to the imaginary part of the effective refractive index of the guided mode and is calculated using the following relation:
\begin{equation}
    \alpha (dB/cm)=8.686\times k_0 Im(n_{eff})\times 10^4
\end{equation}
where $k_0=\frac{2\pi}{\lambda}$ is the number of waves in free space, $\lambda(\mu m)$ is the operating wavelength and $Im(n_{eff})$ is the imaginary part of the effective refractive index. 
\cref{fig:3} presents the confinement loss (CL) spectra of the proposed PCF-SPR biosensor for differentiating between normal and cancerous cells in skin, blood, and adrenal gland tissues. In \cref{fig:3}(a), for skin cells, the normal sample (n = 1.36) exhibits a minimum CL of approximately 3.21 $dB/cm$, while the cancerous counterpart (n = 1.38) shows a significantly higher peak CL of 17.67 $dB/cm$ near 2.2 $\mu m$, indicating strong plasmonic interaction. In \cref{fig:3}(b), for blood cells, the normal cell (n = 1.376) demonstrates a low CL of about 13.2 $dB/cm$, whereas the cancerous cell (n = 1.39) reaches a peak CL of 138 $dB/cm$ around 2.8 $\mu m$, reflecting enhanced sensitivity. Similarly, in \cref{fig:3}(c), for adrenal gland cells, the normal condition (n = 1.381) shows a CL minimum of approximately 7.9 $dB/cm$, while the cancerous cell (n = 1.395) exhibits a pronounced peak CL of 115.3 $dB/cm$ near 3.6 $\mu m$. These significant differences in peak confinement losses between normal and cancerous cells across all tissue types confirm the high sensitivity and diagnostic potential of the proposed PCF-SPR biosensor for effective cancer cell detection.

\begin{table*}[htbp]
\caption{Performance analysis of the proposed SPR-PCF biosensor for detecting normal and cancerous cells across three cell types (Not Optimized)}
\label{tab:Table-3}
\centering
\begin{tabular}{c c c c c c c}
\hline
\thead{Type of \\Cancer} & \thead{Type of \\Cell} & \thead{RI of\\Normal Cell} & \thead{RI of Cancer\\Affected Cell} & \thead{Peak loss\\$(dB/cm)$} & \thead{WS\\$(nm/RIU)$} & \thead{R\\$(RIU)$} \\
\hline
\textbf{Skin} & Basal & 1.360 & 1.380 & 17.67 & 15,000 & $3.13\times10^{-6}$ \\
\hline
\textbf{Blood} & Jurkat & 1.376 & 1.390 & 138 & 53,571 & $1.87\times10^{-6}$ \\
\hline
\textbf{Adrenal} & PC12 & 1.381 & 1.395 & 115.3 & 98,214 & $1.92\times10^{-6}$ \\
\hline
\end{tabular}
\end{table*}
\subsection{Wavelength Sensitivity}\label{subsec2}
Wavelength sensitivity (WS) is a fundamental performance metric for evaluating the efficiency of SPR-based biosensors. It quantifies the shift in the resonance wavelength as a function of changes in the analyte's refractive index (RI). As the RI of the surrounding medium varies, such as when distinguishing between normal and cancerous cells, the resonance condition of the sensor shifts accordingly. This shift enables precise detection and identification of unknown analytes. The wavelength interrogation method is commonly employed to determine WS, which is expressed as:
\begin{equation}
S_\lambda(nm/RIU)=\frac{\Delta\lambda_{peak}}{\Delta n_a}
\end{equation}
where $\Delta\lambda_{peak}$ is the shift of resonant wavelength and $\Delta n_a$ is the change in the refractive index of the analyte. For the proposed SPR-PCF biosensor, the wavelength sensitivity was found to be 15,000 $nm/RIU$ for skin cells, 53,571 $nm/RIU$ for blood cells, and 98,214 $nm/RIU$ for adrenal gland cells. These results demonstrate the sensor’s excellent capability to detect even minor variations in RI, with especially high sensitivity for detecting adrenal gland cancer.


\subsection{Resolution}\label{subsec2}
Resolution is a critical parameter that defines the biosensor’s ability to detect the smallest measurable change in the refractive index (RI) of the analyte. It determines the sensor's precision in distinguishing minute RI differences, which is essential for applications such as early-stage cancer detection where cellular RI changes are subtle. The resolution R is inversely related to the wavelength sensitivity and is calculated using the following equation:
\begin{equation}
    R(RIU)=\Delta n_a\times \frac{\Delta\lambda_{min}}{\Delta\lambda_{peak}}=\frac{\Delta\lambda_{min}}{ S_\lambda}
    \label{eq:7}
\end{equation}
where $\Delta\lambda_{min}=0.1 nm$ indicates the minimum wavelength resolution and $\Delta\lambda_{peak}$ denotes the shift of the resonance peak in the wavelength domain. For the proposed SPR-PCF biosensor, the resolution was found to be $3.13\times10^{-6}$ RIU for skin cells, $1.87\times10^{-6}$ RIU for blood cells, and $1.92\times10^{-6}$ RIU for adrenal gland cells. These results indicate the sensor’s excellent precision in detecting minimal RI changes across various cancer cell types, thereby affirming its potential for high-accuracy, label-free cancer diagnostics.

\subsection{Figure of Merit (FOM)}\label{subsec2}
The figure of merit (FOM) is a key parameter used to evaluate the overall performance and detection accuracy of an SPR-based biosensor. It provides a balance between the sensor’s wavelength sensitivity and the sharpness of the resonance peak, which directly influences its ability to resolve closely spaced refractive index variations. FOM is defined as the ratio of wavelength sensitivity to the full width at half-maximum (FWHM) of the loss spectrum and is calculated using the following equation:
\begin{equation}
    FOM(RIU^{-1})=\frac{S_\lambda}{FWHM}
\end{equation}\\
where FWHM (in nm) represents the spectral width of the resonance curve at half its maximum loss value. A higher FOM indicates a sharper resonance and improved sensing performance, allowing for more accurate differentiation between analytes with similar refractive indices. Therefore, optimizing FOM is essential for enhancing the resolution, reliability, and diagnostic capability.
\begin{figure*}[h]
    \centering
    \subfloat[]{
    \includegraphics[width=0.49\linewidth]{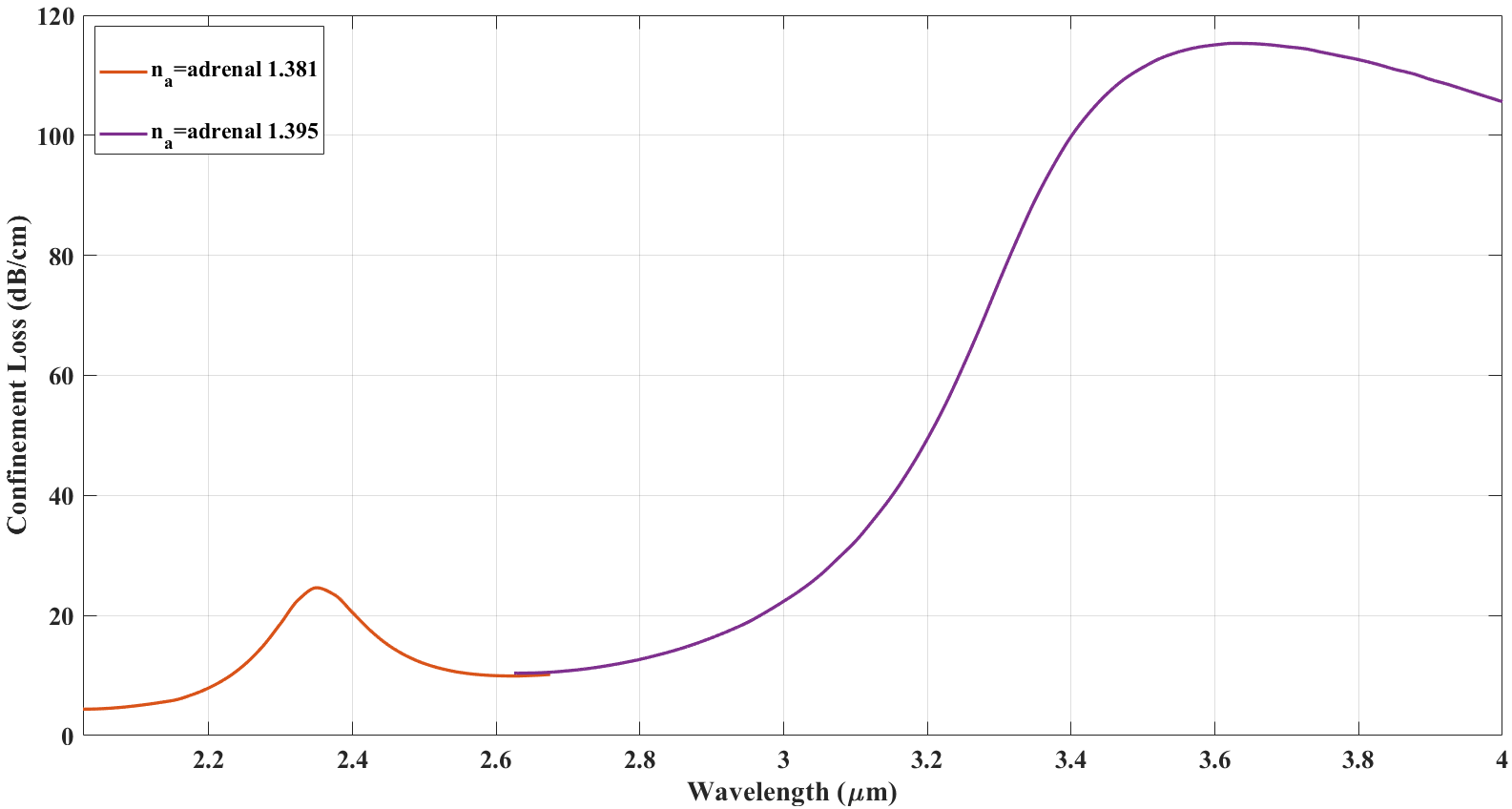}}
    \subfloat[]{
    \includegraphics[width=.49\linewidth]{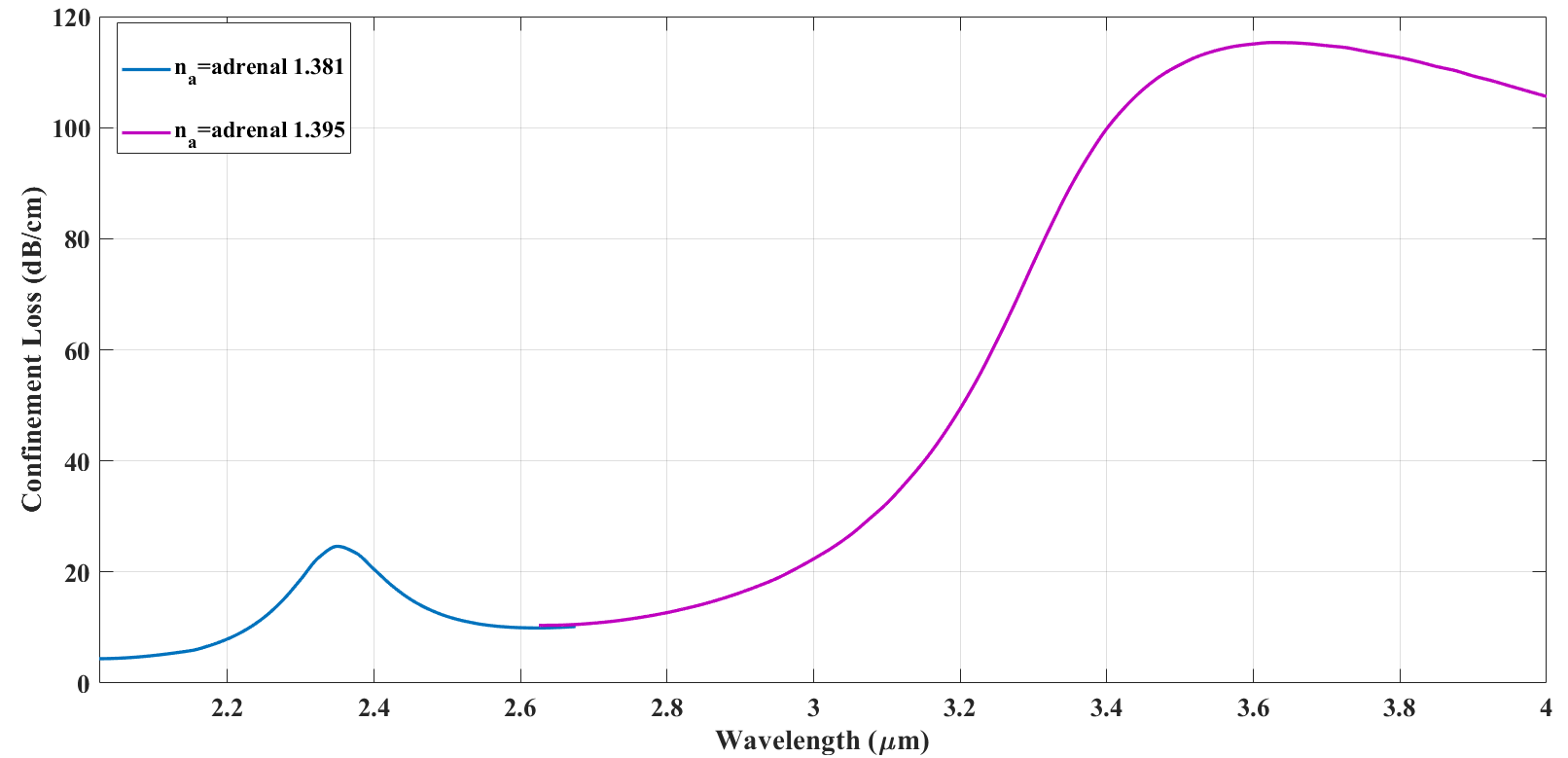}}\\
    \caption{Confinement loss variation of the proposed PCF-SPR biosensor for different air hole radii, all for adrenal gland cell detection: (a) r = 2.5 $\mu m$ and (b) r = 2.7 $\mu m$. The results highlight the influence of air hole geometry on modal confinement and plasmonic interaction strength.
}
    \label{fig:6}
\end{figure*}
\begin{figure*}[h]
    \centering
    \subfloat[]{
    \includegraphics[width=0.49\linewidth]{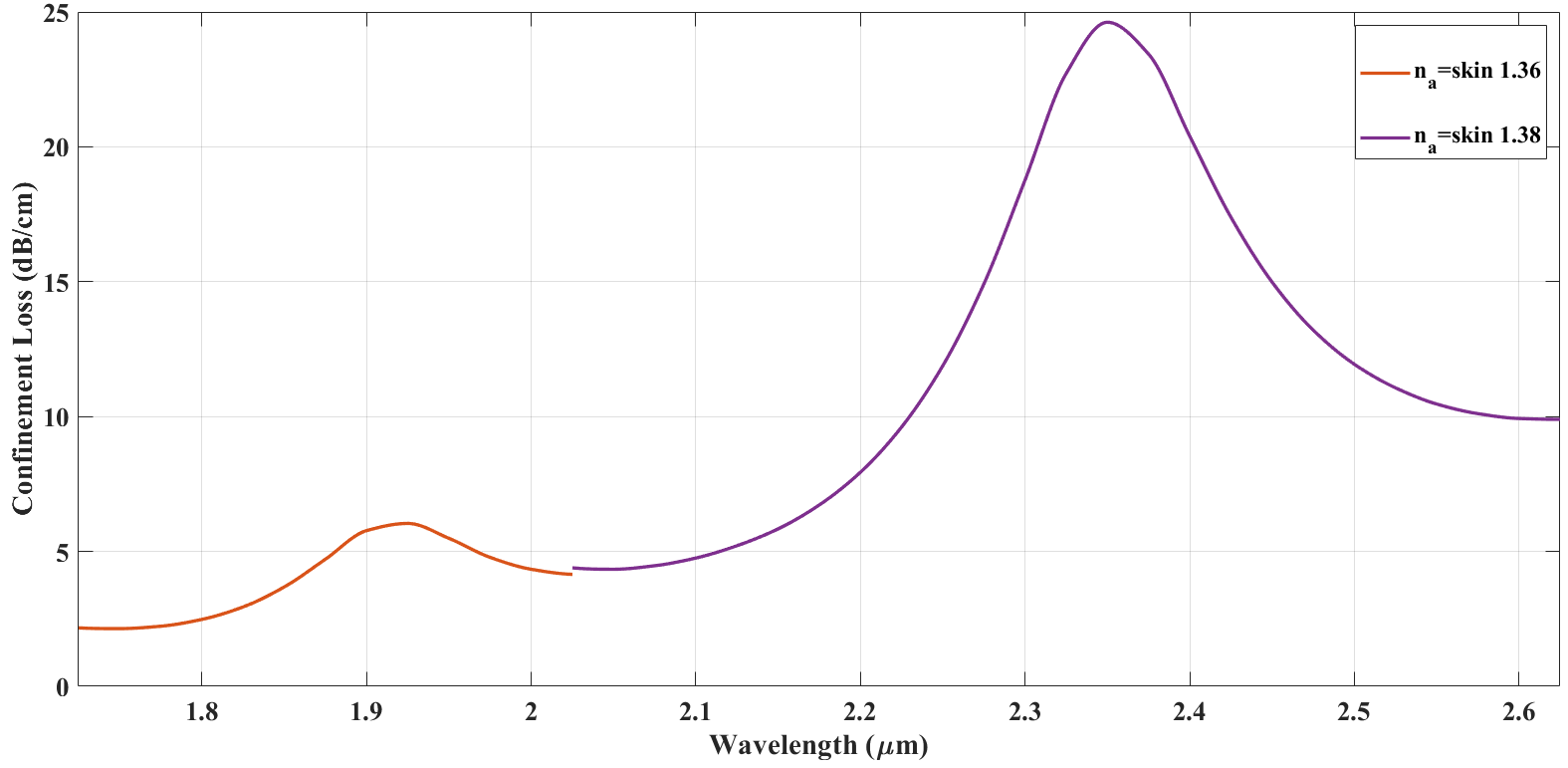}}
    \subfloat[]{
    \includegraphics[width=.49\linewidth]{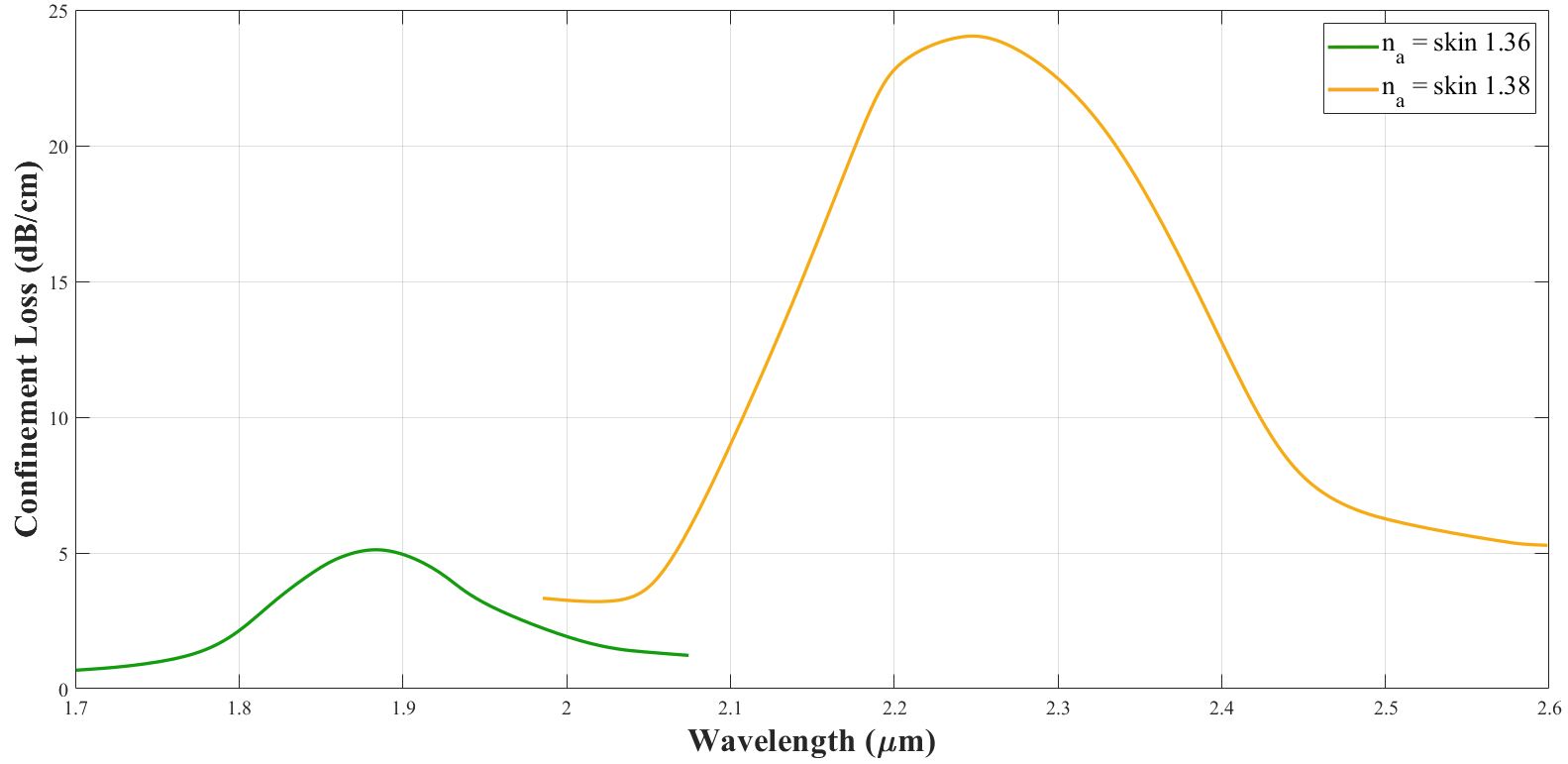}}\\
   
    \caption{Confinement loss variation of the proposed PCF-SPR biosensor for different air hole radii, all for skin cell detection: (a) r = 2.5 $\mu m$ and (b) r = 2.7 $\mu m$. The results highlight the influence of air hole geometry on modal confinement and plasmonic interaction strength.
}
    \label{fig:4}
\end{figure*}
\begin{figure*}[h]
    \centering
    \subfloat[]{
    \includegraphics[width=0.49\linewidth]{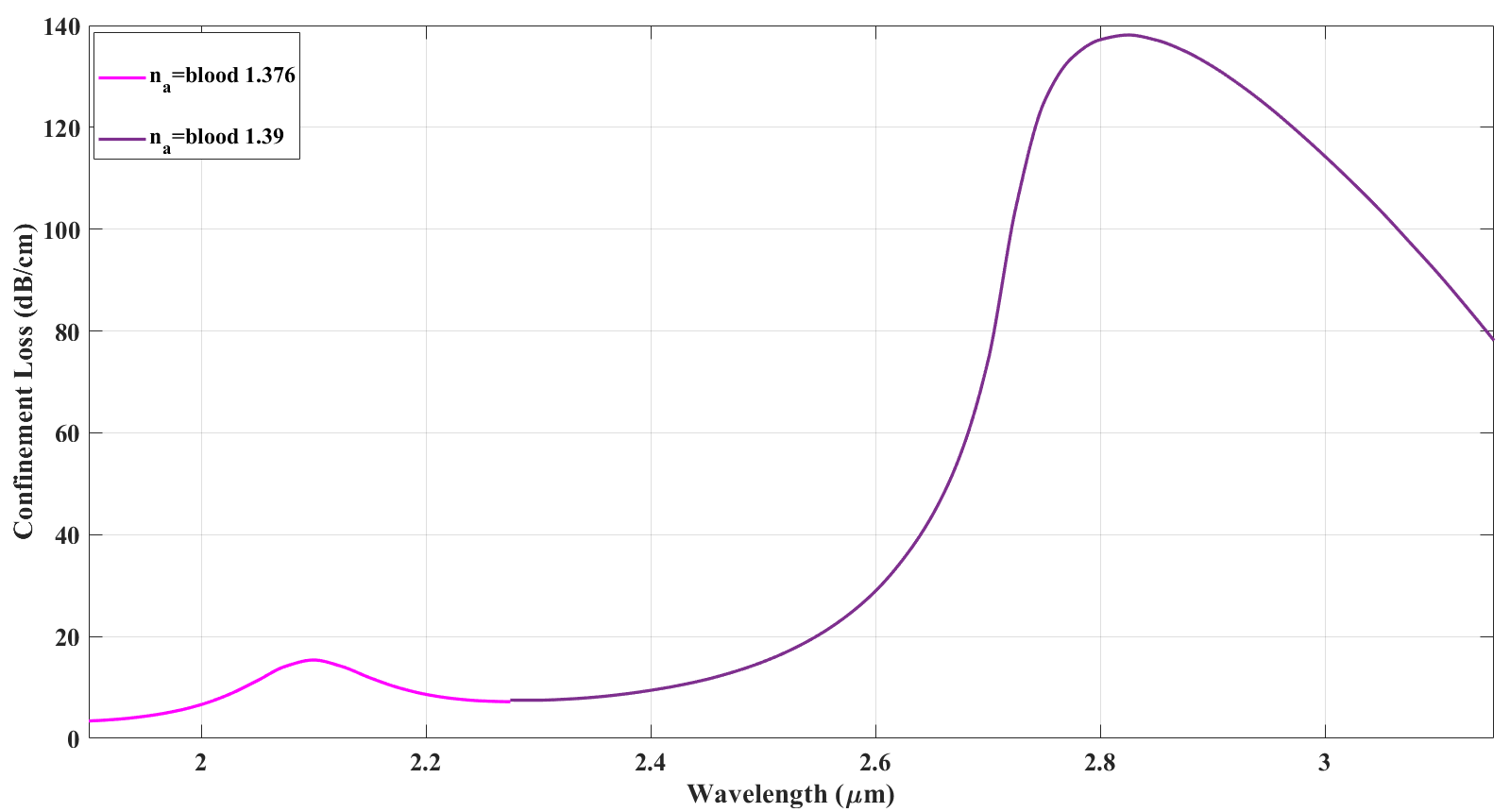}}
    \subfloat[]{
    \includegraphics[width=.49\linewidth]{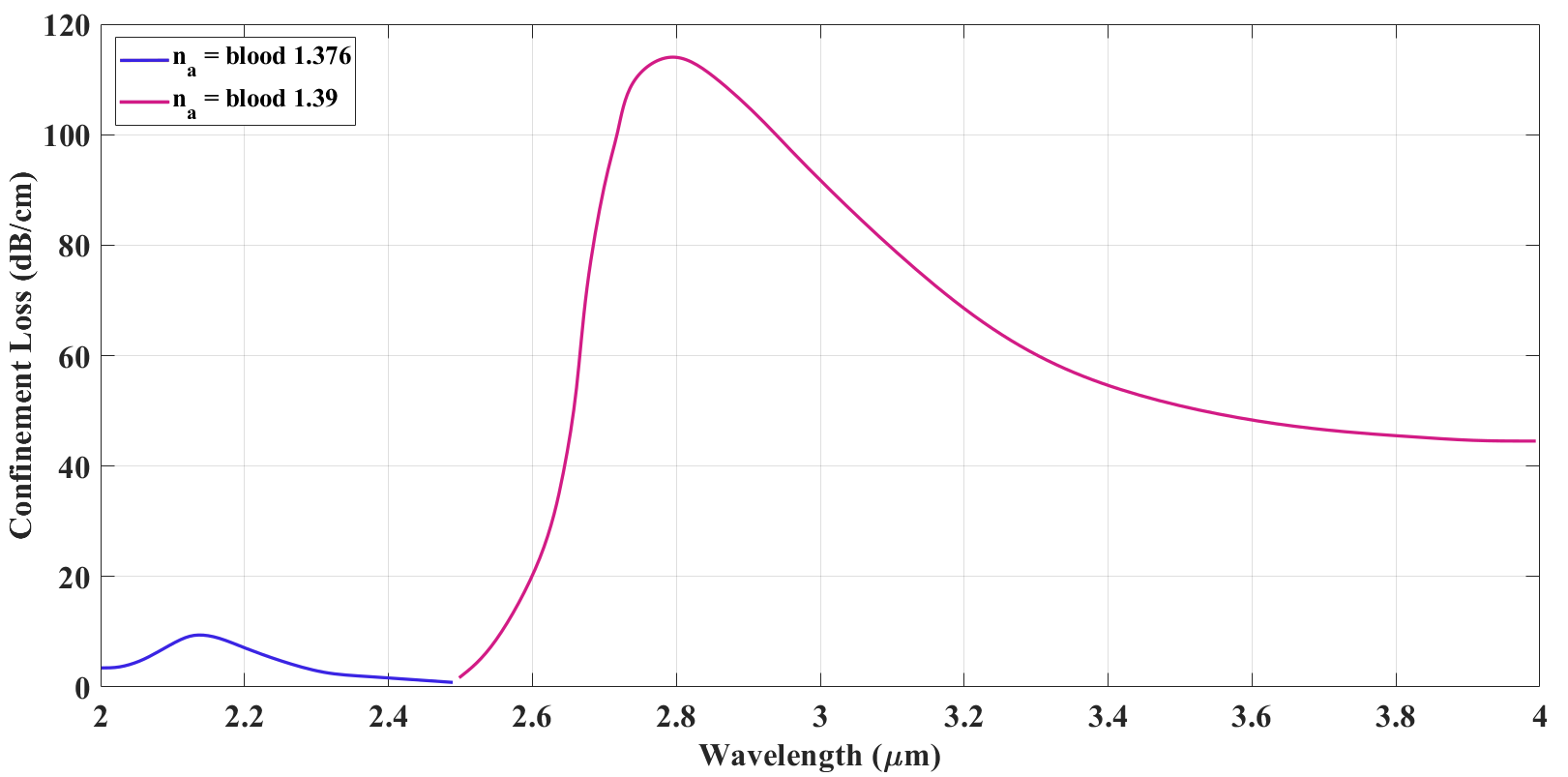}}\\
    \caption{Confinement loss variation of the proposed PCF-SPR biosensor for different air hole radii, all for blood cell detection: (a) r = 2.5 $\mu m$ and (b) r = 2.7 $\mu m$. The results highlight the influence of air hole geometry on modal confinement and plasmonic interaction strength.
}
    \label{fig:5}
\end{figure*}

\begin{table*}[h]
\centering
\caption{Performance Analysis of the Proposed PCF-SPR Biosensor at Air Hole Radius r = 2.5 $\mu m$}
\label{tab:Table-5}
\begin{tabular}{c c c c c c c}
\hline
\thead{Type of\\Cancer} & \thead{Type of\\cell} & \thead{RI} & \thead{Peak loss\\$(dB/cm)$} & \thead{S\textsubscript{$\lambda$}\\$(nm/RIU)$} & \thead{SR\\$(RIU)$} & \thead{FOM\\$(RIU^{-1})$} \\
\hline
\multirow{2}{*}{Skin} & Norm. & 1.36 & 6.03 & \multirow{2}{*}{21,250} &\multirow{2}{*}{$4.71\times10^{-6}$} & \multirow{2}{*}{61.8} \\
& Canc. & 1.38 & 24.62 &  &  &    \\
\hline
\multirow{2}{*}{Blood} & Norm. & 1.376 & 15.39 & \multirow{2}{*}{53,571} & \multirow{2}{*}{$1.87\times10^{-6}$} & \multirow{2}{*}{204.47} \\
& Canc. & 1.39 & 138 &  &  &   \\
\hline
\multirow{2}{*}{Adrenal } & Norm. & 1.381 & 24.61 & \multirow{2}{*}{103,571}  & \multirow{2}{*}{$9.66\times10^{-7}$} & \multirow{2}{*}{306.424} \\
& Canc. & 1.395 & 115.3 &  &  &    \\
\hline
\end{tabular}
\label{tab:3}
\end{table*}

\begin{table*}[htbp]
\centering
\caption{Performance Analysis of the Proposed PCF-SPR Biosensor at Air Hole Radius r = 2.7 $\mu m$}
\label{tab:Table-5}
\begin{tabular}{c c c c c c c}
\hline
\thead{Type of\\Cancer} & \thead{Type of\\cell} & \thead{RI} & \thead{Peak loss\\$(dB/cm)$} & \thead{S\textsubscript{$\lambda$}\\$(nm/RIU)$} & \thead{SR\\$(RIU)$} & \thead{FOM\\$(RIU^{-1})$} \\
\hline
\multirow{2}{*}{Skin} & Norm. & 1.36 & 5.02 & \multirow{2}{*}{18,140} & \multirow{2}{*}{$5.60\times10^{-6}$} & \multirow{2}{*}{58.01} \\
& Canc. & 1.38 & 24 &  &  &   \\
\hline
\multirow{2}{*}{Blood} & Norm. & 1.376 & 8 & \multirow{2}{*}{45,000} & \multirow{2}{*}{$1.38\times10^{-6}$} & \multirow{2}{*}{196} \\
& Canc. & 1.39 & 115.16 &  &  &   \\
\hline
\multirow{2}{*}{Adrenal } & Norm. & 1.381 & 24.61 & \multirow{2}{*}{103,571} & \multirow{2}{*}{$9.57\times10^{-7}$} & \multirow{2}{*}{303.45} \\
& Canc. & 1.395 & 115.3 &  &  &    \\
\hline
\end{tabular}
\label{tab:4}
\end{table*}

\section{Structural Optimization}\label{sec4}
To further enhance the performance of the proposed SPR-PCF biosensor, careful selection and tuning of key structural parameters are essential. Parameters such as the air hole diameter, pitch and metal layer thickness significantly influence the resonance conditions, confinement loss, and sensitivity. This optimization process plays a crucial role in achieving high sensitivity, better resolution, and overall superior biosensing performance.

\begin{figure*}[h]
    \centering
    \subfloat[]{
    \includegraphics[width=0.49\linewidth]{Picture24.png}}
    \subfloat[]{
    \includegraphics[width=0.49\linewidth]{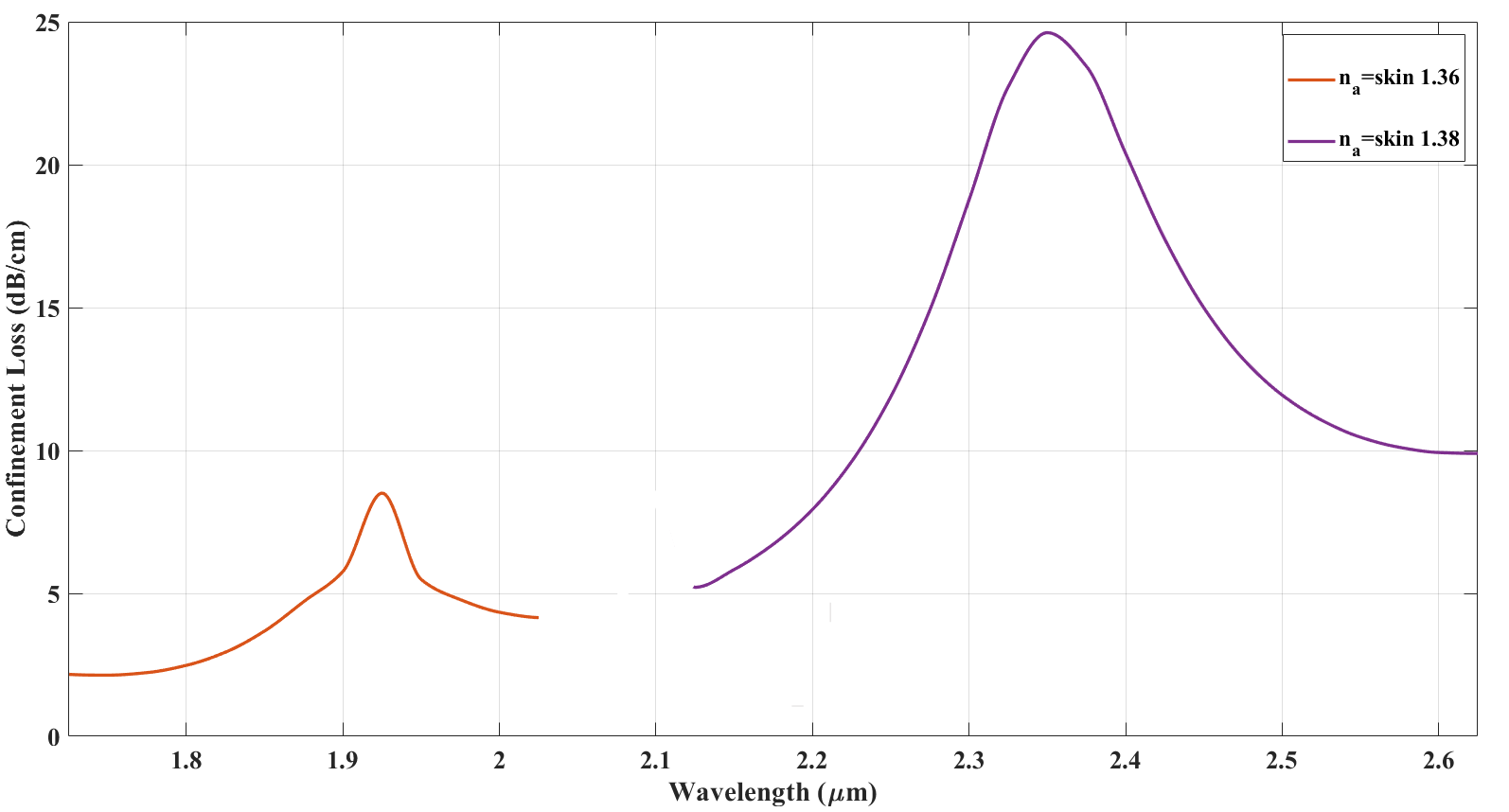}}
    \\[2ex]
    \subfloat[]{
        \includegraphics[width=0.49\linewidth]{Picture28.png}}
        \subfloat[]{
    \includegraphics[width=0.49\linewidth]{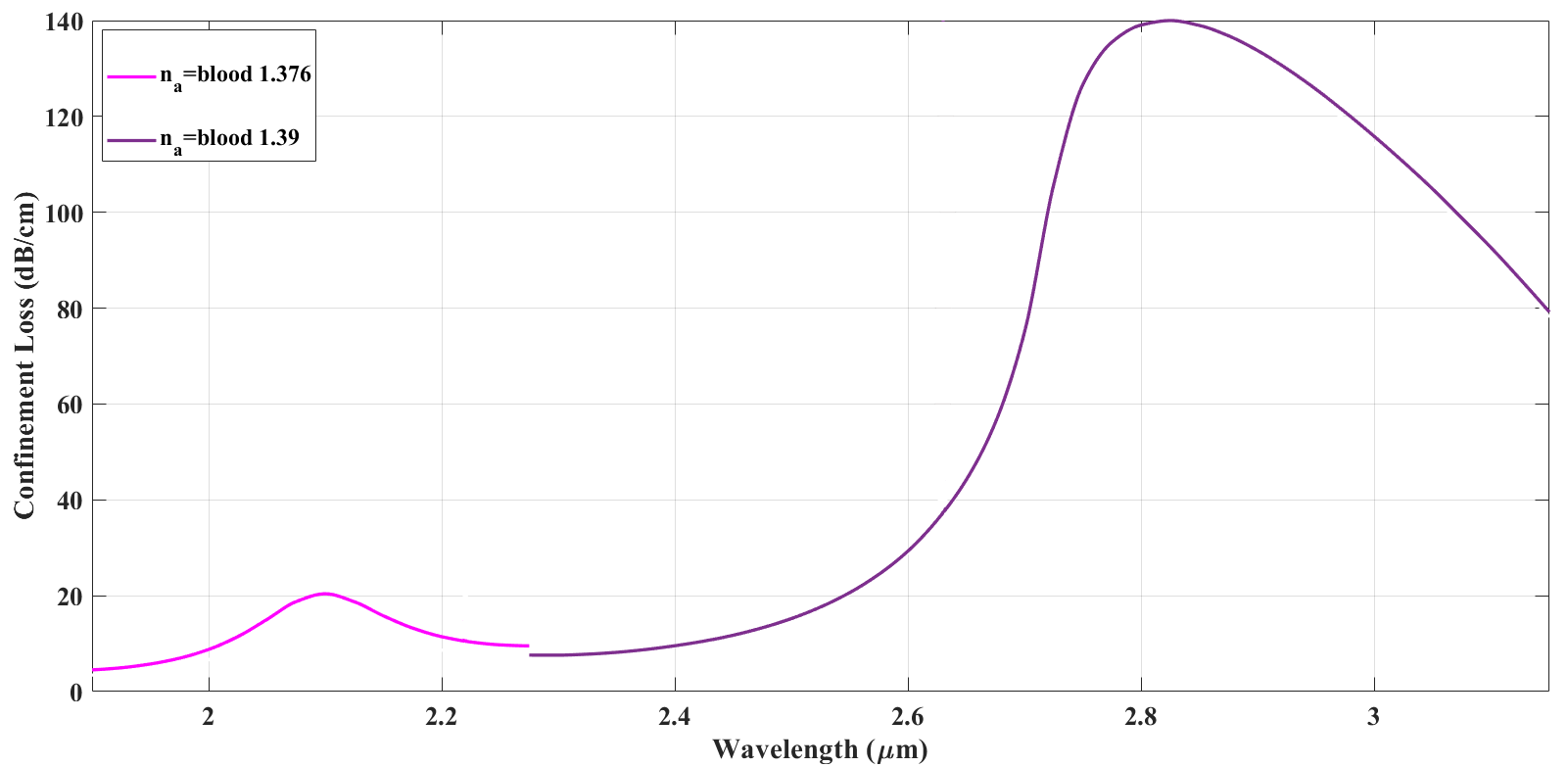}}
    \\[2ex]
     \subfloat[]{
    \includegraphics[width=0.49\linewidth]{Picture32.png}}
    \subfloat[]{
    \includegraphics[width=0.49\linewidth]{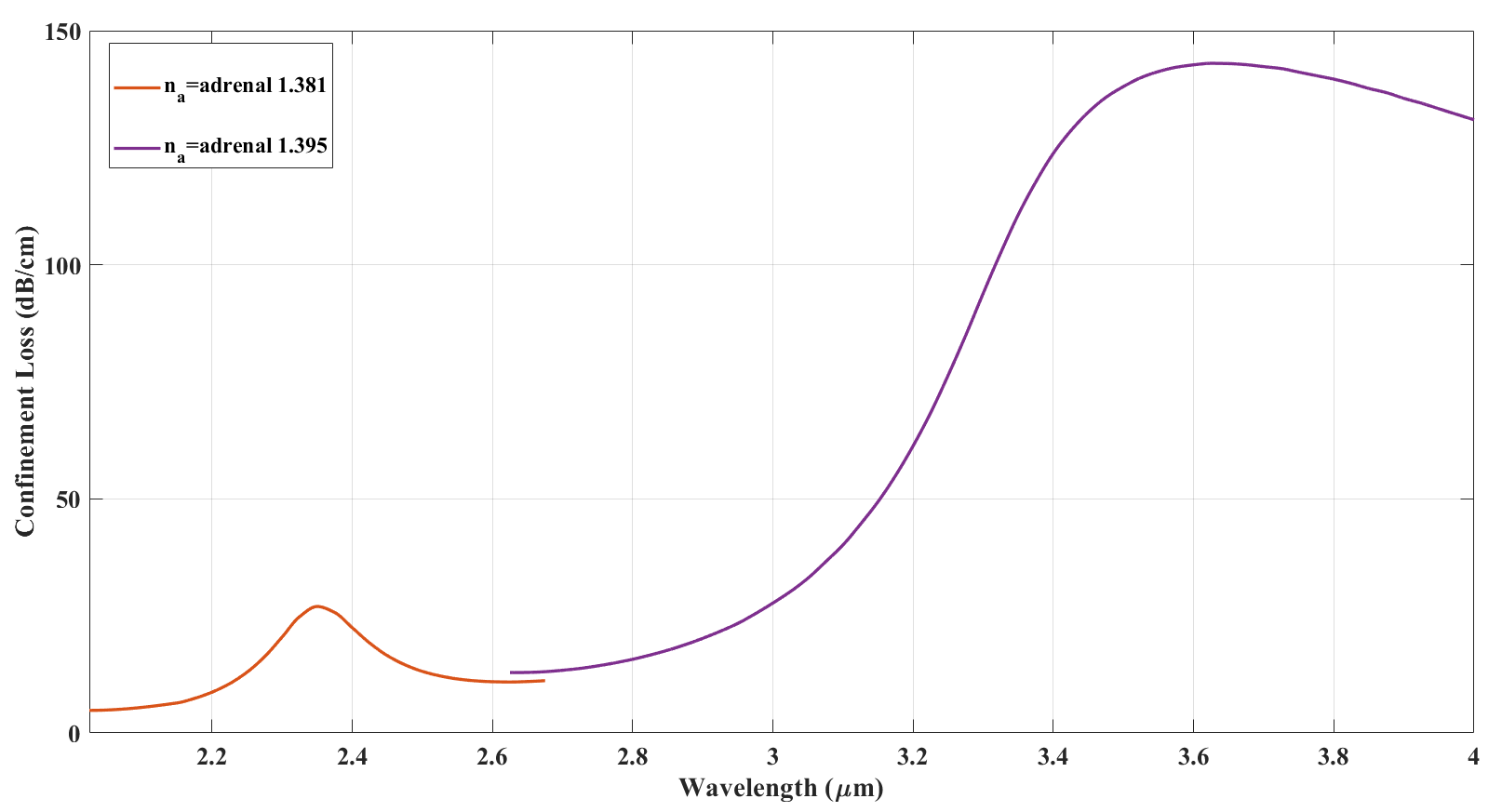}}
    
    \caption{Confinement loss characteristics of the proposed SPR-PCF biosensor for different cell types at two pitch values ($\Lambda$ = 6.0 $\mu m$ and $\Lambda$ = 6.2 $\mu m$): (a–b) skin cells, (c–d) blood cells, and (e–f) adrenal gland cells. In all cases, an increase in pitch results in higher peak confinement loss, while a smaller pitch enhances amplitude sensitivity. The optimal pitch of $\Lambda$ = 6.0 $\mu m$ consistently provides the best trade-off between sensitivity and loss across all three cell types, confirming its suitability for multi-type cancer cell detection.}
    \label{fig:7}
\end{figure*}

\subsection{Airhole Optimization}
Air holes in photonic crystal fiber (PCF) biosensors play a crucial role in controlling the propagation characteristics of the core-guided mode and facilitating efficient coupling with the surface plasmon polariton (SPP) mode at the metal-dielectric interface. The size and arrangement of these air holes directly influence the effective refractive index, confinement of the optical field, and overall sensitivity of the sensor.

By optimizing the air hole radius (r), it is possible to finely tune the modal distribution and enhance phase-matching conditions, thereby improving plasmonic resonance and sensing performance. This section investigates the impact of varying air hole dimensions on the sensor’s performance to identify the optimal configuration that ensures maximum structural efficiency and sensing accuracy.

The simulation was conducted for r = 2.5 $\mu m$ and r = 2.7 $\mu m$ to investigate the effect of air hole radius on the performance of the proposed PCF-SPR biosensor.

 \subsubsection{\textbf{For skin cell}}
 The results, presented in \cref{fig:4} and summarized in \cref{tab:3} and \cref{tab:4}, provide insight into the relationship between air hole geometry and confinement loss. As shown in \cref{fig:4}(a) and \cref{fig:4}(b), the peak confinement loss (CL) decreases as the air hole radius increases from 2.5 $\mu m$ to 2.7 $\mu m$. This behavior is attributed to the larger air holes enabling stronger confinement of the core-guided mode, thereby minimizing the leakage of optical energy into the cladding. The increase in air hole radius enhances the refractive index contrast between the core and cladding regions, promoting tighter mode confinement and reducing radiative losses. However, despite this reduction in CL at r = 2.7 $\mu m$, the overall best sensing performance, considering both sensitivity and plasmonic resonance efficiency, was achieved at r = 2.5 $\mu m$, which provides an optimal balance between modal confinement and SPR interaction strength for effective skin cancer cell detection.


\subsubsection{\textbf{For blood cell}}
The results presented in \cref{fig:5} and summarized in \cref{tab:3} and \cref{tab:4} correspond to the air hole optimization analysis for blood cell detection using the proposed PCF-SPR biosensor. As illustrated in \cref{fig:5}(a) and \cref{fig:5}(b), the peak confinement loss (CL) decreases systematically as the air hole radius (r) increases from 2.5 $\mu m$ to 2.7 $\mu m$. This decreasing trend is attributed to enhanced confinement of the core-guided mode due to the larger air holes, which effectively reduce optical leakage into the cladding. A larger hole radius increases the index contrast between the core and cladding regions, leading to stronger modal confinement and lower radiative losses. Although the reduction in confinement loss at r = 2.7 $\mu m$ indicates improved light guidance, the best overall sensor performance, considering both optimized sensitivity and effective plasmonic coupling, was observed at r = 2.5 $\mu m$. This radius provides an optimal balance between field confinement and surface plasmon resonance interaction, making it the most favorable configuration for detecting blood cancer cells in this design.


\subsubsection{\textbf{For adrenal gland cell}}
The results shown in \cref{fig:6} and detailed in \cref{tab:3} and \cref{tab:4} pertain to the air hole radius optimization for adrenal gland cell detection using the proposed PCF-SPR biosensor. As illustrated in \cref{fig:6}(a) and \cref{fig:6}(b), the peak confinement loss (CL) decreases systematically as the air hole radius (r) increases from 2.5 $\mu m$ to 2.7 $\mu m$. This trend is attributed to improved confinement of the core-guided mode enabled by the larger air holes, which effectively minimize light leakage into the cladding by enhancing the refractive index contrast. Although this reduction in confinement loss at r = 2.7 $\mu m$ suggests improved optical guidance, the overall best performance, considering both optimized sensitivity and efficient surface plasmon resonance coupling, was achieved at r = 2.5 $\mu m$. This radius offers an ideal balance between strong field localization and sufficient plasmonic interaction, making it the most effective configuration for detecting adrenal gland cancer cells with high accuracy.


\begin{table*}[htbp]
\centering
\caption{Performance Analysis of the Proposed PCF-SPR Biosensor for $\Lambda=6.0$ $\mu m$}
\label{tab:Table-5}
\begin{tabular}{c c c c c c c }
\hline
\thead{Type of\\Cancer} & \thead{Type of\\cell} & \thead{RI} & \thead{Peak loss\\$(dB/cm)$} & \thead{S\textsubscript{$\lambda$}\\$(nm/RIU)$} & \thead{SR\\$(RIU)$} & \thead{FOM\\$(RIU^{-1})$} \\
\hline
\multirow{2}{*}{Skin} & Norm. & 1.36 & 6.03 & \multirow{2}{*}{21,250} & \multirow{2}{*}{$4.71\times10^{-6}$} & \multirow{2}{*}{61.8} \\
& Canc. & 1.38 & 24.62 &  &  &   \\
\hline
\multirow{2}{*}{Blood} & Norm. & 1.376 & 15.39 & \multirow{2}{*}{53,571} & \multirow{2}{*}{$1.87\times10^{-6}$} & \multirow{2}{*}{204.47} \\
& Canc. & 1.39 & 138 &  &  &  \\
\hline
\multirow{2}{*}{Adrenal } & Norm. & 1.381 & 24.61 & \multirow{2}{*}{103,571} & \multirow{2}{*}{$9.66\times10^{-7}$} & \multirow{2}{*}{306.424} \\
& Canc. & 1.395 & 115.3 &  &  &   \\
\hline
\end{tabular}
\label{tab:5}
\end{table*}
\begin{table*}[htbp]
\centering
\caption{Performance Analysis of the Proposed PCF-SPR Biosensor for $\Lambda=6.2$ $\mu m$}
\label{tab:Table-5}
\begin{tabular}{c c c c c c c }
\hline
\thead{Type of\\Cancer} & \thead{Type of\\cell} & \thead{RI} & \thead{Peak loss\\$(dB/cm)$} & \thead{S\textsubscript{$\lambda$}\\$(nm/RIU)$} & \thead{SR\\$(RIU)$} & \thead{FOM\\$(RIU^{-1})$} \\
\hline
\multirow{2}{*}{Skin} & Norm. & 1.36 & 8.5 & \multirow{2}{*}{21,000}& \multirow{2}{*}{$5.01\times10^{-6}$} & \multirow{2}{*}{108} \\
& Canc. & 1.38 & 24.7 &  &  &   \\
\hline
\multirow{2}{*}{Blood} & Norm. & 1.376 & 20 & \multirow{2}{*}{52,142} & \multirow{2}{*}{$1.68\times10^{-6}$} & \multirow{2}{*}{157} \\
& Canc. & 1.39 & 140 &  &  &    \\
\hline
\multirow{2}{*}{Adrenal } & Norm. & 1.381 & 25.54 & \multirow{2}{*}{91,428} & \multirow{2}{*}{$1.36\times10^{-6}$} & \multirow{2}{*}{285.26} \\
& Canc. & 1.395 & 147.48 &  &  &    \\
\hline
\end{tabular}
\label{tab:6}
\end{table*}

\begin{table*}[htbp]
\centering
\caption{Performance analysis of the proposed SPR-PCF biosensor for $t_{Au}=40$ $nm$}
\label{tab:Table-5}
\begin{tabular}{c c c c c c c }
\hline
\thead{Type of\\Cancer} & \thead{Type of\\cell} & \thead{RI} & \thead{Peak loss\\$(dB/cm)$} & \thead{S\textsubscript{$\lambda$}\\$(nm/RIU)$} & \thead{SR\\$(RIU)$} & \thead{FOM\\$(RIU^{-1})$} \\
\hline
\multirow{2}{*}{Skin} & Norm. & 1.36 & 6.03 & \multirow{2}{*}{21,250} & \multirow{2}{*}{$4.71\times10^{-6}$} & \multirow{2}{*}{61.8} \\
& Canc. & 1.38 & 24.62 &  &  &    \\
\hline
\multirow{2}{*}{Blood} & Norm. & 1.376 & 15.39 & \multirow{2}{*}{53,571} & \multirow{2}{*}{$1.87\times10^{-6}$} & \multirow{2}{*}{204.47} \\
& Canc. & 1.39 & 138 &  &  &   \\
\hline
\multirow{2}{*}{Adrenal } & Norm. & 1.381 & 24.61 & \multirow{2}{*}{103,571} & \multirow{2}{*}{$9.66\times10^{-7}$} & \multirow{2}{*}{306.424} \\
& Canc. & 1.395 & 115.3 &  &  &    \\
\hline
\end{tabular}
\label{tab:7}
\end{table*}
\begin{table*}[htbp]
\centering
\caption{Performance analysis of the proposed SPR-PCF biosensor for $t_{Au}=42$ $nm$}
\label{tab:Table-5}
\begin{tabular}{c c c c c c c }
\hline
\thead{Type of\\Cancer} & \thead{Type of\\cell} & \thead{RI} & \thead{Peak loss\\$(dB/cm)$} & \thead{S\textsubscript{A}\\($RIU^{-1})$} & \thead{SR\\$(RIU)$} & \thead{FOM\\$(RIU^{-1})$} \\
\hline
\multirow{2}{*}{Skin} & Norm. & 1.36 & 6.0 & \multirow{2}{*}{21,000}  & \multirow{2}{*}{$9.54\times10^{-6}$} & \multirow{2}{*}{99} \\
& Canc. & 1.38 & 14.7 &  &  &    \\
\hline
\multirow{2}{*}{Blood} & Norm. & 1.376 & 30 & \multirow{2}{*}{64,285}& \multirow{2}{*}{$1.02\times10^{-6}$} & \multirow{2}{*}{206} \\
& Canc. & 1.39 & 152 &  &  &    \\
\hline
\multirow{2}{*}{Adrenal } & Norm. & 1.381 & 23 & \multirow{2}{*}{85,714}  & \multirow{2}{*}{$4.98\times10^{-6}$} & \multirow{2}{*}{243.24} \\
& Canc. & 1.395 & 117.88 &  &  &    \\
\hline
\end{tabular}
\label{tab:8}
\end{table*}

\begin{figure*}[h]
    \centering
    \subfloat[]{
    \includegraphics[width=0.49\linewidth]{Picture24.png}}
    \subfloat[]{
    \includegraphics[width=0.49\linewidth]{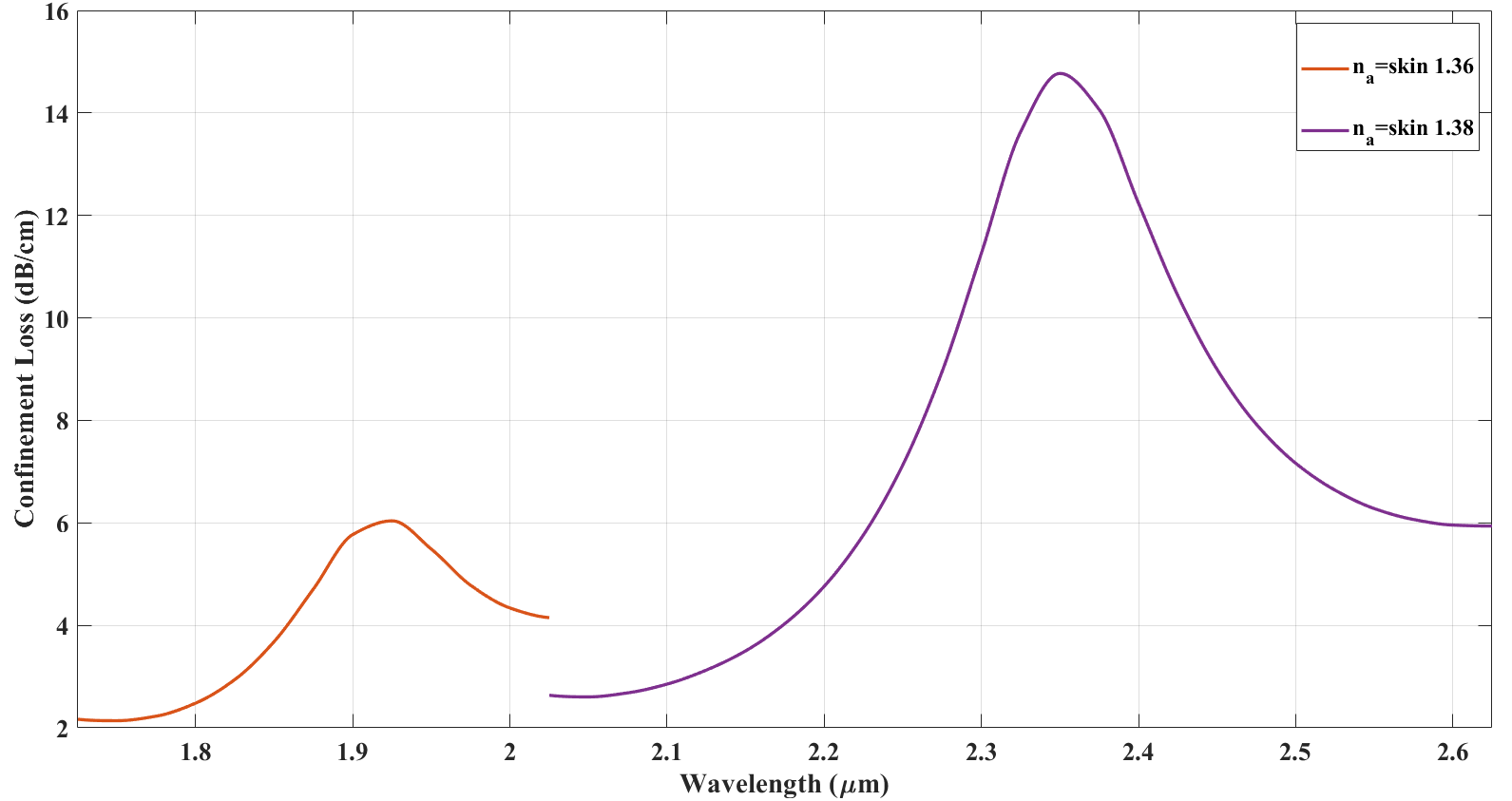}}
    \\
    \subfloat[]{
        \includegraphics[width=0.49\linewidth]{Picture28.png}}
        \subfloat[]{
    \includegraphics[width=0.49\linewidth]{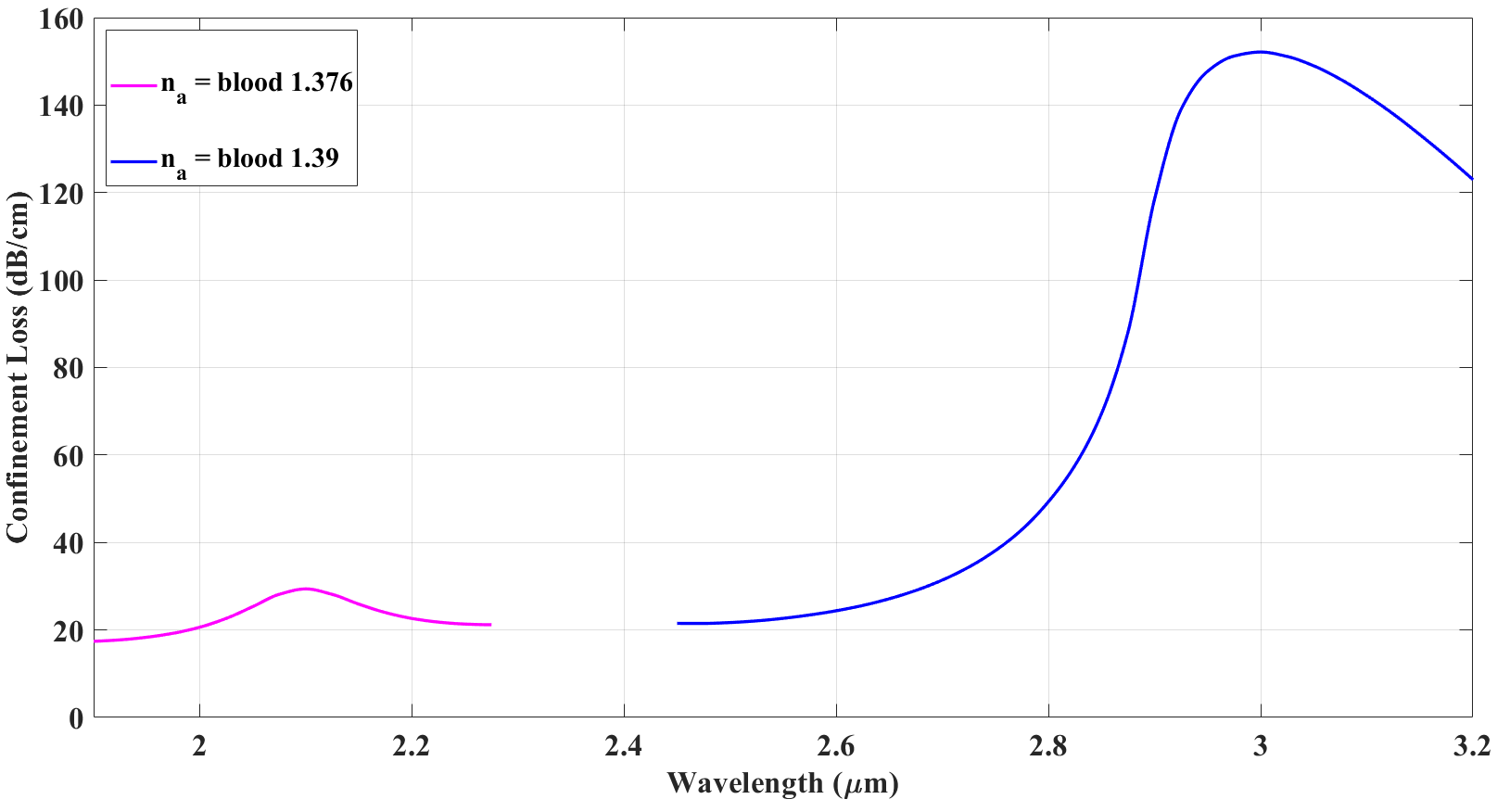}}
    \\
     \subfloat[]{
    \includegraphics[width=0.49\linewidth]{Picture32.png}}
    \subfloat[]{
    \includegraphics[width=0.49\linewidth]{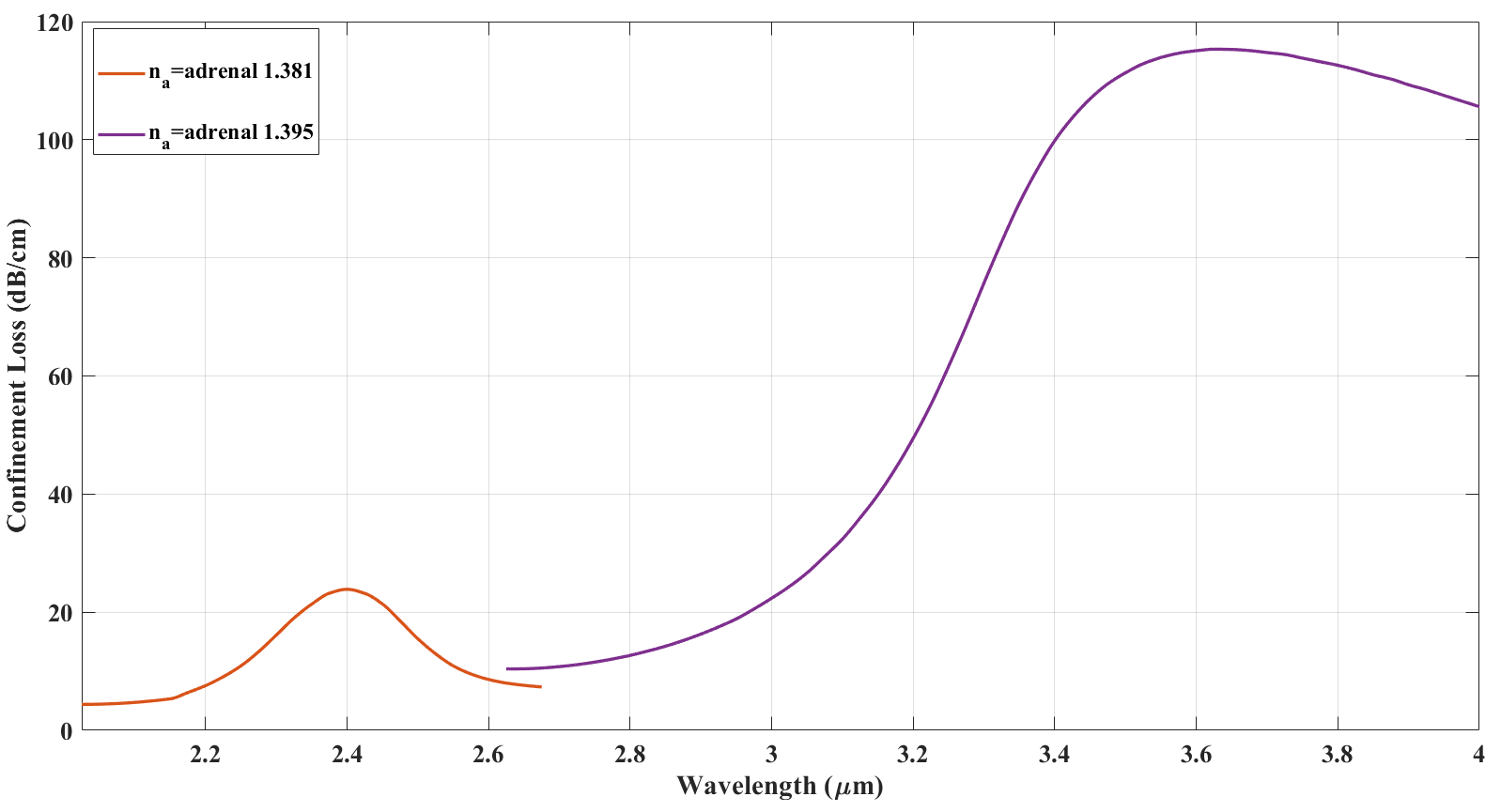}}
    
    \caption{Confinement loss characteristics of the proposed SPR-PCF biosensor for three different cell types at two gold layer thicknesses ($(t_{Au})$ = 40 $nm$ and $(t_{Au})$ = 42 $nm$): (a–b) skin cells, (c–d) blood cells, and (e–f) adrenal gland cells. In all cases, a gold thickness of 40 $nm$ results in improved performance, including higher wavelength sensitivity, better figure of merit, and enhanced resolution. Based on these results, $(t_{Au})$ = 40 $nm$ is identified as the optimal thickness for effective multi-type cancer cell detection.}
    \label{fig:8}
\end{figure*}
\begin{table*}[h!]
\centering
\caption{Performance comparison of the proposed SPR-PCF biosensor (optimized) with recently reported biosensors}
\label{tab:Table-11}
\begin{tabular}{c c c c c c }
\hline
\thead{Ref.} & \thead{Year} & \thead{Structure} & \thead{RI range} & \thead{Max. WS\\$(nm/RIU)$} & \thead{R\\$(RIU)$}  \\
\hline
\cite{mittal2023spiral} & 2023 &  Spiral Shaped & 1.36-1.401 & ---  & $2.33\times10^{-4}$  \\
\hline
\cite{abdelghaffar2023highly} & 2023 & V-shaped & --- & 6,214 & $1.66\times10^{-5}$  \\
\hline
\cite{jain2024pcf} & 2024 & PCF based & 1.36-1.392 & 6,250 & ---\\
\hline
\cite{nagavel2024highly} & 2024 & D-shaped & 1.392-1.399 & 12,857  & --- \\
\hline
\textbf{Proposed} & 2025 & SPR-PCF & 1.36-1.395 & 103,571 &$9.57\times10^{-7}$  \\
\hline

\hline
\end{tabular}
\label{tab:9}
\end{table*}
\subsection{Pitch $(\Lambda)$Optimization}
The simulation was performed for two pitch values, $(\Lambda)$ = 6.0 $\mu m$ and $(\Lambda)$ = 6.2 $\mu m$, to evaluate their effect on the confinement loss and amplitude sensitivity of the proposed SPR-PCF biosensor. For skin cells, the confinement loss characteristics shown in \cref{fig:7}(a) and \cref{fig:7}(b) indicate that an increase in pitch $(\Lambda)$ from 6.0 $\mu m$ to 6.2 $\mu m$ results in higher peak confinement loss, while a smaller pitch enhances the amplitude sensitivity. This trend reflects a trade-off between loss and sensitivity, with $\Lambda$ = 6.0 $\mu m$ providing an optimal balance. A similar behavior is observed in the case of blood cells, as depicted in \cref{fig:7}(c) and \cref{fig:7}(d). Increasing the pitch leads to elevated confinement losses, while reducing the pitch improves amplitude sensitivity. The data in \cref{tab:5} and \cref{tab:6} confirm that $\Lambda$  = 6.0 $\mu m$ offers the most favorable performance metrics for blood cell detection as well. For adrenal gland cells, \cref{fig:7}(e) and \cref{fig:7}(f) further reinforce this trend. As the pitch increases, the confinement loss rises, whereas a reduced pitch enhances the amplitude sensitivity. Again, the optimal pitch is found to be 6.0 $\mu m$, striking a practical compromise between maximizing sensitivity and minimizing loss. This establishes $\Lambda$ = 6.0 $\mu m$ as a robust design parameter for multi-type cancer cell detection using the proposed SPR-PCF biosensor, making it a preferred choice



\subsection{ Optimizing Au Layer Thickness $(t_{Au})$}
The effect of gold (Au) layer thickness on the performance of the SPR-PCF biosensor was analyzed by simulating two thicknesses: 40 $nm$ and 42 $nm$. Key performance metrics including wavelength sensitivity (WS), figure of merit (FOM), and resolution were evaluated for skin, blood, and adrenal gland cells, with results shown in \cref{fig:8}(a)–\cref{fig:8}(f) and \cref{tab:7} and \cref{tab:8}. A gold thickness of 40 $nm$ consistently demonstrated superior performance for both skin and adrenal gland cells, yielding higher WS, improved FOM, and better resolution. 
While the performance difference for blood cells was less pronounced, the 40 $nm$ layer still offered slightly better sensitivity and resolution.
\section{Conclusion}
In conclusion, the proposed SPR-PCF biosensor offers a significant advancement in the field of optical cancer diagnostics by integrating photonic crystal fiber with surface plasmon resonance to enable highly sensitive and real-time detection of cancerous cells, including basal cells (skin cancer), Jurkat cells (blood cancer), and adrenocortical cells (adrenal cancer). The biosensor's structural design, incorporating a silica core, circular air holes, a gold sensing layer, and a $V_2O_5$ adhesion layer, enhances plasmonic coupling and field confinement, resulting in improved sensing performance. Through systematic optimization of key parameters such as air hole radius, pitch, and gold layer thickness, the sensor achieves outstanding metrics, including a maximum wavelength sensitivity of $1035.71$ $nm/RIU$, a figure of merit (FOM) of 306.424 $ RIU^{-1}$, and a best resolution of $9.57 \times10^{-7} RIU$. These results demonstrate the biosensor's strong potential for early-stage cancer detection with high resolution and selectivity. Its consistent performance across diverse cell types confirms its robustness and versatility, making it a promising candidate for future medical diagnostics and environmental biosensing applications. This work highlights the practical relevance and transformative potential of SPR-PCF biosensors in improving diagnostic accuracy and advancing accessible cancer detection technologies.

\section{Acknowledgements}
The authors would like to express their sincere gratitude to the Institute for Theoretical and Computational Physics Research for providing the necessary facilities, computational resources, and a supportive research environment that made this work possible. Their valuable support is gratefully acknowledged.

\section{Author Contributions}
T.M. conducted the literature review and identified the research gap that laid the foundation for the study. He also contributed to the initial design of the device. M.N.N.A. and T.M. collaboratively performed the device simulations. Data analysis was carried out by T.M. and M.N.N.A., while N.S.L. prepared all the figures presented in the paper. The initial manuscript was drafted by T.M. and M.N.N.A., with Z.T.N. providing critical review and revisions. All authors contributed to the finalization of the manuscript and approved its submission.

\section{Declarations}
The authors declare that they have no competing interests or conflicts of interest related to this work.

\section{Funding declaration}
This research received no external funding and was carried out as a collaborative effort among the authors.

\bibliographystyle{IEEEtran}
\bibliography{citation}

\vfill

\end{document}